\documentclass[twocolumn, prl, superscriptaddress,notitlepage]{revtex4-1}
\usepackage{graphicx}
\usepackage{physics}
\usepackage{epsfig}
\usepackage{color}
\usepackage{xspace}
\usepackage{array}
\usepackage{amsmath}
\usepackage{amsfonts}
\usepackage{ulem}
\usepackage[dvipsnames]{xcolor}
\usepackage{colortbl}
\usepackage[breaklinks,colorlinks,linkcolor=blue,citecolor=blue,urlcolor=blue]{hyperref}

\allowdisplaybreaks

\begin{document}
\title{Quench Dynamics of Thermal Bose Gases Across Wide and Narrow Feshbach}

\author{Xiaoyi Yang}
\affiliation{MOE Key Laboratory for Nonequilibrium Synthesis and Modulation of Condensed Matter,
Shaanxi Province Key Laboratory of Quantum Information and Quantum Optoelectronic Devices, School of Physics,
Xi'an Jiaotong University, Xi'an 710049, China}

\author{Ren Zhang}
\email{renzhang@xjtu.edu.cn}
\affiliation{MOE Key Laboratory for Nonequilibrium Synthesis and Modulation of Condensed Matter,
Shaanxi Province Key Laboratory of Quantum Information and Quantum Optoelectronic Devices, School of Physics,
Xi'an Jiaotong University, Xi'an 710049, China}

\begin{abstract}
Using high-temperature virial expansion, we study the quench dynamics of the thermal Bose gases near a wide, narrow, and intermediate Feshbach resonance. Our results show that the shallow bound state near Feshbach resonance leads to interesting phenomena. Near the wide Feshbach resonance, the long-time $\hat{n}_{\bf k}$ oscillates when the scattering length $a_{s}$ is quenched from zero to large but with finite positive values. The oscillation frequency $\omega=E_{\rm b}/\hbar$ with $E_{\rm b}$ being the binding energy. When $a_{s}$ is quenched to infinity or negative value, the oscillation vanishes. Near the narrow Feshbach resonance, the interaction should be characterized by a two-channel model. When the background scattering length $a_{\rm bg}\gtrsim\lambda$, there is an oscillation in the long-time dynamics, and the frequency is determined by the energy of the shallow bound state in the open channel. When $a_{\rm bg}<0$ or $0<a_{\rm bg}\ll\lambda$, there is no shallow bound state in the open channel, hence no long-time oscillation. We check our conclusion using some realistic systems, and the results are consistent with our conclusion.
\end{abstract}
\maketitle

{\it Introduction-} Thanks to Feshbach resonances, the pairwise interaction between atoms can be controlled flexibly by tuning external fields, and the equilibrium properties in strongly interacting atomic gases have been intensively studied \cite{FR1,FR2}. The timescale of the Hamiltonian manipulation can be much smaller than the relaxation time. As such, ultracold atomic gases have also become one of the most ideal platforms to investigate non-equilibrium physics, including the quench dynamics~\cite{dynamics1,dynamics2,dynamics3,dynamics4,Bosequenchexp,dynamics5,dynamics6}.

Here, quench dynamics refers to the evolution of initial states under an abruptly changed Hamiltonian. For instance, the $s$-wave scattering length $a_{s}$ can be controlled by the magnetic field. When $a_{s}$ is modulated by time-dependent external fields, the particle number on the non-condensation mode will exponentially grow, i,e. Bose-Einstein condensation (BEC) is depleted. If the modulation phase is suddenly changed by $\pi$, it was found that the excited particle number decreases, i.e. BEC revives \cite{ChengChindynamics}. Motivated by this phenomenon, a new kind of echo theory has also been raised in BEC, which can be realized by quenching $a_{s}$ or the trapping potential \cite{SU11echo,MBE}. In the same spirit, there are many other studies on quench dynamics via quenching parameters of the Hamiltonian. Many-body localization and thermalization can be distinguished by the quench dynamics of the entanglement entropy~\cite{MBLdynamics}. The topology of Hamiltonian of band insulators can be extracted in the quench dynamics of linking number~\cite{topo1,topo2,topo3}. The dynamical fractal has been established in quantum gases with discrete scaling symmetry~\cite{zheyudynamics}.

In the seminal experiment by Cambridge group~\cite{Bosequenchexp}, a series of universal quench dynamics of Bose gas have been revealed by quenching the interaction from zero to unitary. Both degenerate and thermal Bose gases composited of $^{39}$K are studied near a Feshbach resonance located at $\sim 402.7G$. This is a resonance of intermediate width, $s_{\rm res}\sim2.1$~\cite{FR1}. In the follow-up theoretical studies, it has been treated as a wide one, and the comparison of theoretical and experimental results are satisfactory for both degenerate and thermal gases \cite{Chaodynamics,SunPRL}. A natural question arises: what are the  effects induced by resonance width? To address this question, we focus on the dynamics of thermal Bose gas near a Feshbach resonance with varying width.

The virial expansion builds a connection between the few-body and the many-body physics \cite{Virial1,Virial2,Virial3,Virial4,Virial5,Virial6,Virial7,Virial8,Virial9,Virial10,Virial11,Virial12,Virial13,Virial14}. It works well when comparing with experiments \cite{Virialexp1,Virialexp2,Virialexp3,Virialexp4,Virialexp5,Virialexp6,Virialexp7,Virialexp8}. The control parameter is the fugacity $z=e^{\mu/(k_{B}T)}$, where $\mu$ is the chemical potential and $k_{B}$ is the Boltzmann constant. At high temperature, $\mu$ is large and negative. Therefore, $z<1$. This method has been applied to equilibrium quantum gases near wide Feshbach resonances \cite{wideF_virial,wideF_virial1,wideF_virial2,wideF_virial3,wideF_virial4} and narrow Feshbach resonances \cite{Virial5,narrowF_virial,narrowF_virial1,narrowF_virial2,narrowF_virial3,narrowF_virial4}. 
Recently, it has also been implemented to the quench dynamics of Bose gas near a wide Feshbach resonance \cite{SunPRL}. As such, it is natural to generalize the virial expansion for dynamics to the narrow Feshbach resonance, and a crossover from a wide one to a narrow one. Our main results are summarized in Table \ref{table1}.

\begin{table}
\caption{\centering Long-time behavior of the momentum distribution $n_{\bf k}$ dynamics near Feshbach resonance.\label{table1}}
\centering
\begin{tabular}{|p{2.5cm}<{\centering}|p{2.5cm}<{\centering}|p{2.5cm}<{\centering}|}
\hline
Resonance width&Oscillation&Non-oscillation\\
\hline
$s_{\rm res}\gg1$ (wide resonance )& $s$-wave scattering length $a_{s}>0$&$s$-wave scattering length $a_{s}<0$ or $a_{s}=\infty$\\
\hline
$s_{\rm res}\ll1$ (narrow resonance) or $s_{\rm res}\sim1$ (intermediate width)&Background scattering length $a_{\rm bg}\gtrsim\lambda$&Background scattering length $a_{\rm bg}<0$ or $0<a_{\rm bg}\ll\lambda$\\
\hline
\end{tabular}
\end{table}

{\it Virial expansion for dynamic-}
Let us first review the basics of virial expansion for quench dynamics. We consider an equilibrium thermal Bose gas with temperature $T$, then quench
the interaction from zero to finite or unitary. The Hamiltonian becomes $\hat{H}=\hat{H}_{0}+\hat{V}$ with $\hat{V}$ denoting the interaction.
For later time $t>0$, the system evolve under the full Hamiltonian $\hat{H}$, and eventually achieve a new equilibrium state. Some universal physics can be revealed  in prethermal process.
To this end, we could measure an observable ${\hat W}$,  the expectation value ${\cal W}(t)$ of which  can be written as \cite{SunPRL}
\begin{align}
\label{Wt}
\mathcal{W}(t)=\frac{\mathrm{Tr}[e^{- \beta (\hat{H}_0 -\mu \hat{N})} e^{i\hat{H}t} \hat{W} e^{-i \hat{H}t}]}{\mathrm{Tr}[e^{-\beta (\hat{H}_0 - \mu \hat{N})}]},
\end{align}
where $\beta=1/{k_BT}$ denotes the inverse temperature, $\hat{N}$ is the total particle number of Bose gas. Here and after forth, we set $\hbar=k_B=1$ for convenience. The exact evaluation of ${\cal W}(t)$ is formidable in a many-body system because $\hat{W}$ does not commute with the Hamiltonian.

At high temperature, we expand the observable ${\mathcal W}(t)$, instead of the thermodynamic potential $\Omega$, in terms of  the fugacity $z$. 
Up to the order of $z^2$, $\mathcal W(t)$ is expressed as
\begin{align}
\mathcal{W}(t)=X_1z+\left(-Q_1X_1+X_2\right)z^2+{\cal O}\left(z^3\right). \label{eq:WExpressTo2}
\end{align}
Here, $X_n$ and $Q_n$ are defined as 
\begin{align}
X_n &=\mathrm{Tr}_n [ \Theta (t) e^{- \beta \hat{H}_0} e^{it \hat{H}} \hat{W} e^{-it \hat{H}}] \notag\\
&=\underset{\alpha,\beta,\gamma}\sum e^{-\beta E_\alpha ^{(n)}}G_{\beta \alpha} ^{(n)*}(t)\langle \psi_{\beta} ^{(n)}|\hat{W}|\psi_\gamma ^{(n)}\rangle G_{\gamma\alpha} ^{(n)}(t),\\
Q_n &=\mathrm{Tr}_n [ e^{- \beta \hat{H}_0} ],
\end{align}
respectively. 
$n=1,2\cdots$ indicates the particle number, 
$E^{(n)}_\alpha$ and $\psi^{(n)}_\alpha$ represents the energy and wave function of $n$-particle non-interaction state, respectively. $\Theta(t)$ is the step function. $G^{(n)}(t)$ is retarded Green's function of $n$-particle interacting system, and it is defined as
\begin{align}
G_{\gamma\alpha} ^{(n)}(t) &= \langle \psi_\gamma^{(n)}  | \Theta (t) e^{-it \hat{H}} | \psi_\alpha^{(n)} \rangle \notag\\
&=\frac{i}{2\pi}\int_{-\infty} ^{\infty}d\omega e^{-i\omega t}G_{\gamma\alpha} ^{(n)}(\omega +i0^+).
\end{align}
As a result, by solving the $n$-particle problem, the evolution of the many-body system can be obtained, and the accuracy can be improved by increasing $n$.

The same as the experiment, 
we consider the dynamics of the particle number in the ${\bf k}$-mode, i.e., $\hat W=\hat n_{\mathbf k}$. For a single particle system, $\hat H=\hat H_0$, and
\begin{align}
X_1=\mathrm{Tr}_1 \left[\Theta (t) e^{-\beta \hat H_0}e^{it\hat H_0}\hat n_{\mathbf k} e^{-it \hat H_0}\right] =e^{- \beta\mathbf k^2/(2m)}.
\end{align}
Therefore, $X_1$ and $Q_n$ is independent on time $t$, The evolution of momentum distribution $\delta n_{\mathbf k}=n_{\mathbf k}(t)-n_{\mathbf k}(0)$ only depends on $X_2$ in Eq.(\ref{eq:WExpressTo2}). Specifically,
\begin{align}
\label{deltan}
\delta n_{\mathbf k}=\left[X_2(t)-X_2(0)\right]z^2,
\end{align}
which can be obtained by only solving the two-body problem. For the two-body problem, the non-interacting wave function is labeled by $| \psi _a^{(2)} \rangle =|{\bf P},{\bf  q}\rangle$ with ${\bf P}$ and ${\bf q}$ being the total momentum and the relative momentum of two bosons, respectively. The corresponding energy reads $E_a^{(2)} =  P^2/(4m)+q^2/m$, where $m$ is the reduced mass. According to the Lippman-Schwinger function, the retarded Green's function can be written as
\begin{align}
G_{\alpha \beta }^{(2)}(s) &= G_{\alpha \beta }^{(0)}(s) + G_{\alpha \beta }^{(0)}(s)T_2(s)G_{\alpha \beta }^{(0)}(s) \notag \\
&= \left[\frac{\langle \mathbf{q_\alpha}|\mathbf{q_\beta}\rangle}{s-\varepsilon _{\mathbf{q_\alpha}}}+\frac{T_2(s)}{(s-\varepsilon _{\mathbf{q_\alpha}})(s-\varepsilon _{\mathbf{q_\beta}})}\right]\delta_{\mathbf{P_\alpha,P_\beta}},\label{eq:green-single}
\end{align}
where $s=\omega+i0^+$, $\varepsilon _{\mathbf{q_\alpha}}=\mathbf q^2_\alpha/m$ is the kinetic energy of the relative motion, $T_{2}(s)$ denotes the T-matrix of the two-body scattering. We have also used the free Green's function for the relative motion $G_0(s)=(s-\mathbf q^2/m)^{-1}$ in the second line of Eq.(\ref{eq:green-single}).

{\it Wide and narrow resonances-}
Before embarking on the difference between wide and narrow resonance, let us recall the two-body scattering theory. The generic relation between scattering amplitude $f({\bf k}'\leftarrow {\bf k})$ and scattering T-matrix $T({\bf k}', {\bf k};E)$ is $f({\bf k}'\leftarrow {\bf k})=-\frac{m}{4\pi}T_{2}({\bf k}', {\bf k};E)$. For the partial wave scattering, the scattering amplitude $f_{\ell}(k)$ is defined by the partial wave scattering matrix $s_{\ell}=e^{2i\delta_{\ell}}$,
\begin{align}
f_{\ell}(k)=\frac{s_{\ell}-1}{2ik}=-\frac{1}{ik-k/\tan\delta_{\ell}(k)},
\end{align}
where $\delta_{\ell}(k)$ is the energy-dependent phase shift of the $\ell$-th partial wave. In this manuscript, we consider the $s$-wave scattering, and the corresponding  T-matrix is then written as
\begin{align}
\label{T2}
T_{2}(s)=\frac{4\pi/m}{-\sqrt{ms}/\tan\delta_{0}-\sqrt{-ms}}
\end{align}
In the effective field theory, $\sqrt{ms}/\tan\delta_{0}=-1/a_{s}+r_{\rm eff}ms/2+\cdots$ with $a_{s}$ and $r_{\rm eff}$ denoting the $s$-wave scattering length and the effective range, respectively. For wide resonance, the effective range effect can be ignored, and $a_{s}$ is the only parameter to characterize the pairwise interaction. For narrow resonance, one has to include the effective range to incorporate the energy-dependent of phase shift. For the van der Waals interaction between two atoms, one has to resort to the complicated quantum defect theory to obtain the exact phase shift \cite{QDT1,QDT2}. To be simple, we consider a two-channel square model to mimic the interaction between atoms. In the basis spanned by closed and open channel, the interaction can be written as
\begin{align}
\label{two-channelmodel}
V(r)=\begin{cases} 
\begin{bmatrix} -V_o & W \\ W & -V_c+\delta\mu_{B} B \end{bmatrix}, & \text{for } r<r_0; \\ \begin{bmatrix} 0 & 0 \\ 0 & \infty \end{bmatrix}, & \text{for } r>r_0,
\end{cases}
\end{align}
where $V_{c}$ and $V_{o}$ represent  the closed channel and open channel potential, and $W$ is the inter-channel coupling strength. $r_{0}$ is the potential range, and the corresponding energy scale is $E_0=1/(mr_0^2)$. $\delta\mu_{B}$ is the magnetic momentum difference between closed and open channel. By tuning the magnetic field, a series of Feshbach resonances appear (supplementary material). Although this toy model does not quantitatively  capture the interaction potential detail, it is an insight model to present the qualitative picture. By solving the two-channel model, the effective phase shift can be analytically obtained. Upon substituting the phase shift into Eq.(\ref{T2}) and (\ref{eq:green-single}), the evolution of momentum distribution $\delta n_{\bf k}(t)$ can be obtained. To precisely distinguish wide and narrow Feshbach resonance, we define the parameter
\begin{align}
s_{\rm res}=\frac{a_{\rm bg}}{r_{0}}\frac{\delta\mu_{B}\Delta}{E_{0}},
\end{align}
where $a_{\rm bg}$ is the background scattering length determined by $V_{o}$. $\Delta$ is the resonance width in the magnetic field aspect, and is determined by the inter-channel coupling $W$. When $s_{\rm res}\gg1$, it is a wide resonance; when $s_{\rm res}\ll1$, it is a narrow resonance; when $s_{\rm res}\sim1$, it is of intermediate width.

\begin{figure}
\centering
\includegraphics[width=0.5\textwidth]{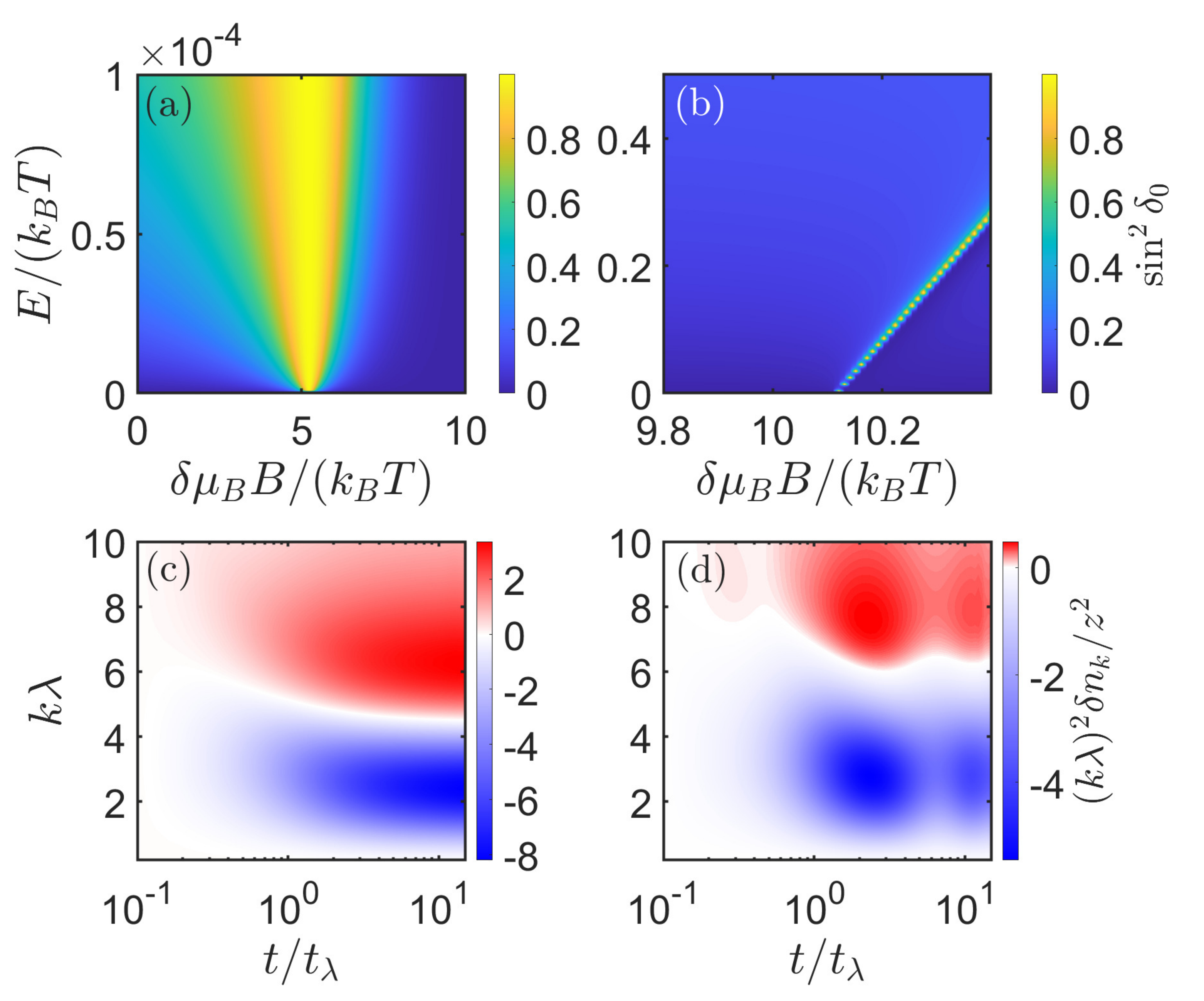}
\caption{The phase shift $\sin^{2}\delta_0$ and momentum distribution $\delta n_{\bf k}$ evolution of thermal Bose gas near wide and narrow resonance. The interaction is quenched by abruptly changing the magnetic field to resonance. $s_{\rm res}\sim260$ for the wide resonance (a) and $s_{\rm res}\sim0.04$ for the narrow resonance (b). (c): Near the wide resonance, the low-momentum $n_{\bf k}$ monotonically decreases and the high-momentum $n_{\bf k}$ monotonically increases. The critical momentum is around $k\lambda=4.5$. (d): Near the narrow resonance, $n_{\bf k}$ shows damped oscillation when it decreases or increases.}
\label{fig1}
\end{figure}

In Fig.~\ref{fig1}, we depict $\sin^{2}\delta_{0}$ as a function of the magnetic field and incident energy for both wide resonance (a) and narrow resonance (b). The $s_{\rm res}\sim260$ and $s_{\rm res}\sim0.04$ for the wide and narrow resonance. When $\delta_{0}=\pi/2$, i.e., $\sin^{2}\delta_{0}=1$, the resonance happens. It is clear that the phase shift of wide resonance almost does not depend on the incident energy, as shown in (a). In contrast, the phase shift of narrow resonance strongly depends on the incident energy, as shown in (b). For both cases, we quench the interaction by abruptly changing the magnetic field to $B_{\rm res}$, the position of resonance, and measure the dynamics of momentum distribution. Near the wide resonance, the low-momentum ($k\lambda<4.5$) $n_{\bf k}$ decreases monotonically after quenching and tends to a stable value after a long time evolution; the high-momentum ($k\lambda>4.5$) $n_{\bf k}$ increases monotonically and tends to a stable value after a long time evolution. There is a critical momentum ($k\lambda=4.5$), where $n_{\bf k}$ goes up and down, and tends to its initial value. This observation is consistent with experimental results \cite{Bosequenchexp} and theoretical results given by the zero-range potential \cite{SunPRL}. Here $\lambda =\sqrt{2\pi/(mT)}$ denotes the thermal de Broglie wavelength, and we  define a time unit $t_\lambda=1/T$. However, near the narrow resonance, the momentum distribution dynamics show very different behavior in contrast to their wide resonance counterpart. Although the tendency of $n_{\bf k}$ for low-momentum and high-momentum remains, there is oscillation with damping amplitude, which means that there must be an intrinsic energy scale near the narrow resonance. The critical momentum shifts slightly.

{\it Zero-range model-} To understand this phenomenon, let us turn to the zero-range model. For the wide resonance, the two-channel model in Eq.(\ref{two-channelmodel}) can be approximated by a zero-range single-channel model. The two-body scattering T-matrix $T_2(s)$ reduces to
\begin{align}
T_2(s)=\frac{4\pi /m}{a_s^{-1}-\sqrt{-ms}}.\label{eq:T2-single}
\end{align}

\begin{figure}
\centering
\includegraphics[width=0.5\textwidth]{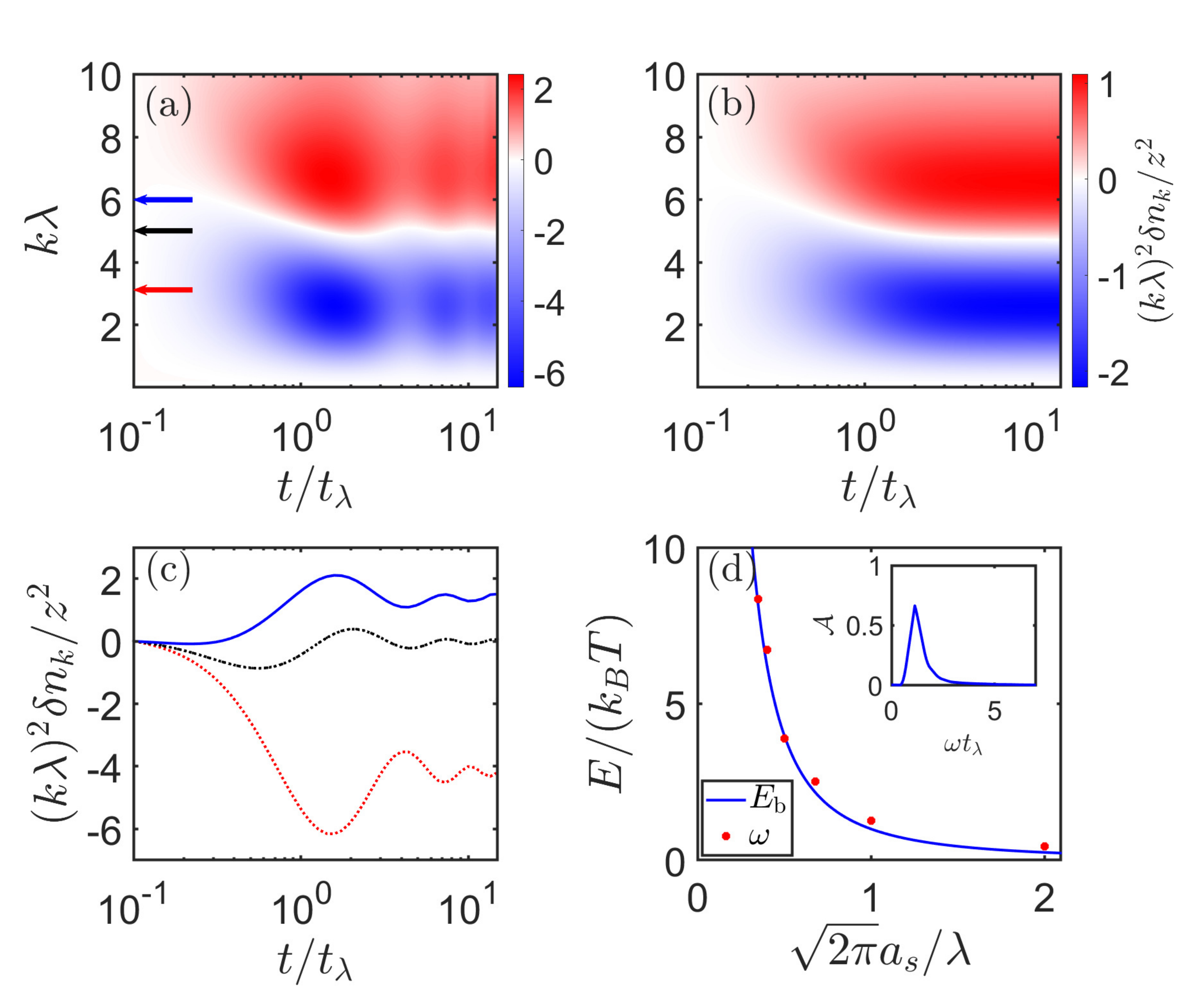}
\caption{The evolution of the momentum distribution when the quenched interaction deviates from the resonance position. (a): Final $a_s=\lambda/\sqrt{2\pi}$. (b): Final $a_s=-\lambda/\sqrt{2\pi}$. (c): $n_{\bf k}$ evolution for three particular momentum; Blue solid line ($k\lambda=6$); Black dashed-dotted (cross momentum); Red dashed line ($k\lambda=3$). (d): The shallow bound state energy $E_{\rm b}=1/(ma_{s}^{2})$ and oscillation frequency $\omega$ of the momentum dynamics. $\omega$ collapses to the bound state energy. Inset: The spectrum ${\cal A}(\omega)$ for the final $a_{s}=\lambda/\sqrt{2\pi}$.}
\label{fig2}
\end{figure}
The pole of $T_{2}(s)$ gives the energy of the shallow bound state, $E_{\rm b}=-1/(ma_{s}^{2})$. When the interaction is quenched to unitary, i.e., $a_{s}=\infty$, the bound state energy vanishes. The Hamiltonian is scale-invariant, and the only relevant length scales are the inter-particle spacing and the thermal de Broglie wavelength. As such, the dynamics driven by the scale-invariant Hamiltonian are universal. Nevertheless, when $a_{s}$ is quenched to a positive finite value, this extra length scale would exhibit itself in the dynamics.

In Fig.~\ref{fig2}, we show the momentum distribution dynamics when the quenched interaction deviates from the resonance position. For final $a_{s}>0$, we find a long-time oscillation in the momentum distribution dynamics, as shown in (a), similar to that near the narrow resonance. The oscillation frequency for different momentum is the same, as depicted in (c). We extract the oscillation frequency for varying $a_{s}$. Our results show that the frequency collapses to the binding energy $|E_{\rm b}|=1/(ma_{s}^{2})$, as shown in (d). Therefore, we conclude that the oscillation in dynamics originates from the shallow bound state. When the final $a_{s}<0$, there is no shallow bound state, hence no oscillation in the dynamics, as shown in (b).

Now we turn to the dynamics near the narrow or intermediate Feshbach resonance, where the single-channel model is not sufficient to characterize the interaction. As such, we need to adopt the two-channel zero range model \cite{zhai_book},
 the two-body scattering T-matrix $T_2$ of which can be written as (supplementary material)
\begin{align}
\label{narrowT2}
T_2(s)=\frac{4\pi/m}{\frac{s-\delta\mu_{B}(B-B_{\rm res})}{a_{\rm bg}\left[s+\delta\mu_{B}\Delta-\delta\mu_{B}(B-B_{\rm res})\right]}-\sqrt{-ms}},
\end{align}
where $B_{\rm res}$ denotes the magnetic field at resonance. The pole of $T_{2}(s)$ in Eq.(\ref{narrowT2}) gives two bound states.
By substituting Eq.(\ref{narrowT2}) into Eq.(\ref{eq:green-single}), we can evaluate the momentum distribution $n_{\bf k}$.

\begin{figure}
\centering
\includegraphics[width=0.35\textwidth]{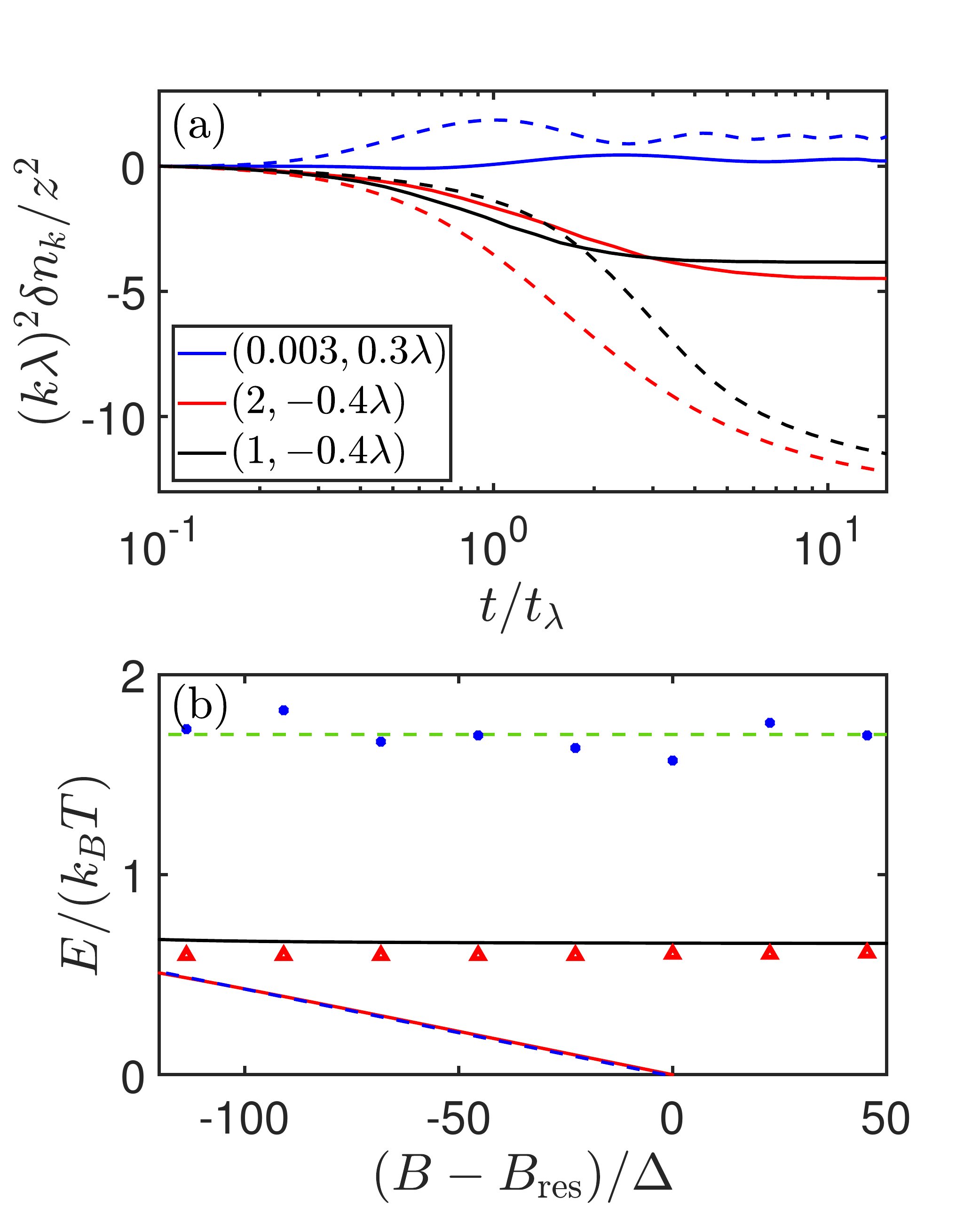}
\caption{(a) Dynamics of $n_{\bf k}$ near the narrow and intermediate Feshbach resonance. The numbers in the legend represent $(s_{\rm res},a_{\rm bg})$. Solid (dashed) curves represent two-channel square (zero-range) model. $k\lambda=7(3)$ for narrow (intermediate) resonance. (b) Binding energy given two-channel square (solid lines) model and zero-range (dashed lines) model. The blue circle and red triangle are the oscillation frequency extracted from the dynamics of $n_{\bf k}$.}
\label{fig3}
\end{figure}

Fig.~\ref{fig3}(a) shows the momentum distribution dynamics of $n_{\bf k}$ after quenching the interaction to resonance. We compare the results of the two-channel square model (solid curves) and zero range model (dashed curves) for some particular momentum. Near the narrow resonance ($s_{\rm res}=0.003$), both results show there is a long-time oscillation, but the frequencies are different. Near resonance of intermediate width ($s_{\rm res}=2$), the dynamics are almost the same as that near the wide resonance. This explains why could the single-channel model give consistent results with the experiment. When $a_{\rm bg}<0$, $n_{\bf k}$ also monotonically decays near the intermediate resonance ($s_{\rm res}=1$), as shown by the black curves. This is because the shallow bound state in open channel is absent in this case.

Near the narrow resonance, the oscillation frequency $\omega$ is determined by the binding energy of the bound state in the open channel instead of the shallow bound state near the threshold. In Fig.~\ref{fig3}(b), we present the binding energy given by the two-channel square model and zero-range model. The binding energy at the threshold of these two models is the same, as illustrated by the blue dashed and red solid lines. However, the other bound states originate from the open channel and  binding energies given by these two models are different, as shown by the green dashed and black solid lines. We extract the oscillation frequency near narrow resonance in Fig.~\ref{fig3}(a). It collapses to the binding energy of the bound state in the open channel, instead of that at the threshold.

\begin{figure}
\centering
\includegraphics[width=0.5\textwidth]{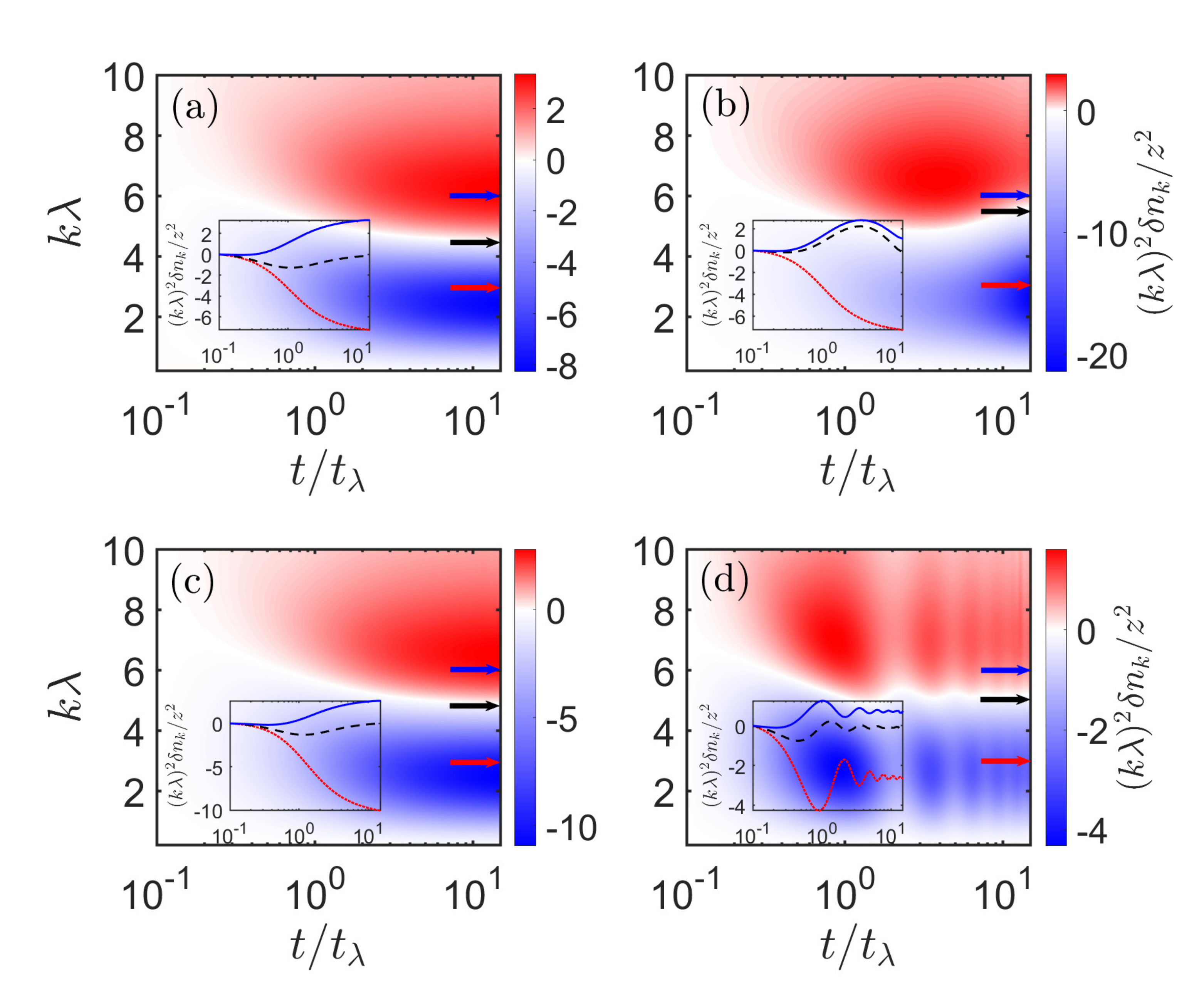}
\caption{The dynamics of $n_{\bf k}$ for realistic systems near Feshbach resonance. (a): $^{133}$Cs near the wide resonance with $s_{\rm res}=560$. (b): $^{133}$Cs near the intermediate resonance with $s_{\rm res}=0.67$ and $a_{\rm bg}=926a_{0}$. (c): $^{7}$Li near the intermediate resonance with $s_{\rm res}=0.8$ and $a_{\rm bg}=-25a_0$. (d): $^{133}$Cs near the narrow resonance with $s_{\rm res}=0.002$ and $a_{\rm bg}=160a_{0}$. $a_{0}$ is the Bohr's radius.}
\label{fig4}
\end{figure}

{\it Application to realistic systems-} We show the quench dynamics of some realistic systems. Here, we consider four different systems including wide, narrow, and intermediate resonance \cite{FR1}. (I): We choose $^{133}\text{Cs}$ near a wide resonance with $s_{\rm res}=560$. When the interaction is quenched to unitary, the evolution of the momentum distribution is as same as that near the wide resonance, as shown in Fig.\ref{fig4} (a). (II): In contrast, for $^{133}\text{Cs}$ near the narrow resonance with $s_{\rm res}=0.002$ and $a_{\rm bg}=160a_{0}$ ($a_{0}$ is the Bohr's radius), shown in Fig.\ref{fig4}(d), we see that the momentum distribution oscillates, which originates from the bound state in the open channel. (III): For quench dynamics near resonance of intermediate width, we choose two systems, $^{133}\text{Cs}$ near resonance with $s_{\rm res}=0.67$ (b) and $^{7}\text{Li}$ near resonance with $s_{\rm res}=0.8$ (c). However, the background scattering length are different for these two systems. For $^{133}\text{Cs}$, $a_{\rm bg}=926a_{0}$ implies a shallow bound state in the open channel, and we see that the momentum distribution oscillates. 
Nevertheless, for $^{7}\text{Li}$, $a_{\rm bg}=-25a_0$, there is no shallow bound state in the open channel, thus no oscillation in the momentum distribution.

In summary, we use virial expansion to study  quench dynamics across the wide and narrow Feshbach resonances. Taking the dynamics of momentum distribution as an example, we show the dynamics can be affected by the bound state, the frequency of oscillation is proportional to the energy of the bound state. Near wide resonance, the relevant bound state is the shallow bound state near the threshold, while near the narrow or intermediate resonance, the relevant bound state is the bound state of the open channel. We check our conclusion using some realistic systems. 

\begin{acknowledgements}
We are grateful to Mingyuan Sun and Xin Chen for helpful discussion.
The work was supported by the National Nature Science Foundation of China (Grant No. 12074307), the National Key R$\&$D Program of China (Grant No. 2018YFA0307601) and the Fundamental Research Funds for the Central Universities (Grant No. 71211819000001).
\end{acknowledgements}


\begin{thebibliography}{54}%
\makeatletter
\providecommand \@ifxundefined [1]{%
 \@ifx{#1\undefined}
}%
\providecommand \@ifnum [1]{%
 \ifnum #1\expandafter \@firstoftwo
 \else \expandafter \@secondoftwo
 \fi
}%
\providecommand \@ifx [1]{%
 \ifx #1\expandafter \@firstoftwo
 \else \expandafter \@secondoftwo
 \fi
}%
\providecommand \natexlab [1]{#1}%
\providecommand \enquote  [1]{#1}%
\providecommand \bibnamefont  [1]{#1}%
\providecommand \bibfnamefont [1]{#1}%
\providecommand \citenamefont [1]{#1}%
\providecommand \href@noop [0]{\@secondoftwo}%
\providecommand \href [0]{\begingroup \@sanitize@url \@href}%
\providecommand \@href[1]{\@@startlink{#1}\@@href}%
\providecommand \@@href[1]{\endgroup#1\@@endlink}%
\providecommand \@sanitize@url [0]{\catcode `\\12\catcode `\$12\catcode
  `\&12\catcode `\#12\catcode `\^12\catcode `\_12\catcode `\%12\relax}%
\providecommand \@@startlink[1]{}%
\providecommand \@@endlink[0]{}%
\providecommand \url  [0]{\begingroup\@sanitize@url \@url }%
\providecommand \@url [1]{\endgroup\@href {#1}{\urlprefix }}%
\providecommand \urlprefix  [0]{URL }%
\providecommand \Eprint [0]{\href }%
\providecommand \doibase [0]{https://dx.doi.org}%
\providecommand \selectlanguage [0]{\@gobble}%
\providecommand \bibinfo  [0]{\@secondoftwo}%
\providecommand \bibfield  [0]{\@secondoftwo}%
\providecommand \translation [1]{[#1]}%
\providecommand \BibitemOpen [0]{}%
\providecommand \bibitemStop [0]{}%
\providecommand \bibitemNoStop [0]{.\EOS\space}%
\providecommand \EOS [0]{\spacefactor3000\relax}%
\providecommand \BibitemShut  [1]{\csname bibitem#1\endcsname}%
\let\auto@bib@innerbib\@empty
\bibitem [{\citenamefont {Chin}\ \emph {et~al.}(2010)\citenamefont {Chin},
  \citenamefont {Grimm}, \citenamefont {Julienne},\ and\ \citenamefont
  {Tiesinga}}]{FR1}%
  \BibitemOpen
  \bibfield  {author} {\bibinfo {author} {\bibfnamefont {C.}~\bibnamefont
  {Chin}}, \bibinfo {author} {\bibfnamefont {R.}~\bibnamefont {Grimm}},
  \bibinfo {author} {\bibfnamefont {P.}~\bibnamefont {Julienne}}, \ and\
  \bibinfo {author} {\bibfnamefont {E.}~\bibnamefont {Tiesinga}},\ }\bibfield
  {title} {\bibinfo {title} {Feshbach resonances in ultracold gases},\ }\href
  {\doibase/10.1103/RevModPhys.82.1225} {\bibfield  {journal} {\bibinfo
  {journal} {Rev. Mod. Phys.}\ }\textbf {\bibinfo {volume} {82}},\ \bibinfo
  {pages} {1225} (\bibinfo {year} {2010})}\BibitemShut {NoStop}%
\bibitem [{\citenamefont {K\"ohler}\ \emph {et~al.}(2006)\citenamefont
  {K\"ohler}, \citenamefont {G\'oral},\ and\ \citenamefont {Julienne}}]{FR2}%
  \BibitemOpen
  \bibfield  {author} {\bibinfo {author} {\bibfnamefont {T.}~\bibnamefont
  {K\"ohler}}, \bibinfo {author} {\bibfnamefont {K.}~\bibnamefont {G\'oral}}, \
  and\ \bibinfo {author} {\bibfnamefont {P.~S.}\ \bibnamefont {Julienne}},\
  }\bibfield  {title} {\bibinfo {title} {Production of cold molecules via
  magnetically tunable feshbach resonances},\ }\href
  {\doibase/10.1103/RevModPhys.78.1311} {\bibfield  {journal} {\bibinfo
  {journal} {Rev. Mod. Phys.}\ }\textbf {\bibinfo {volume} {78}},\ \bibinfo
  {pages} {1311} (\bibinfo {year} {2006})}\BibitemShut {NoStop}%
\bibitem [{\citenamefont {Makotyn}\ \emph {et~al.}(2014)\citenamefont
  {Makotyn}, \citenamefont {Klauss}, \citenamefont {Goldberger}, \citenamefont
  {Cornell},\ and\ \citenamefont {Jin}}]{dynamics1}%
  \BibitemOpen
  \bibfield  {author} {\bibinfo {author} {\bibfnamefont {P.}~\bibnamefont
  {Makotyn}}, \bibinfo {author} {\bibfnamefont {C.~E.}\ \bibnamefont {Klauss}},
  \bibinfo {author} {\bibfnamefont {D.~L.}\ \bibnamefont {Goldberger}},
  \bibinfo {author} {\bibfnamefont {E.~A.}\ \bibnamefont {Cornell}}, \ and\
  \bibinfo {author} {\bibfnamefont {D.~S.}\ \bibnamefont {Jin}},\ }\bibfield
  {title} {\bibinfo {title} {Universal dynamics of a degenerate unitary bose
  gas},\ }\href {\doibase/10.1038/nphys2850} {\bibfield  {journal} {\bibinfo
  {journal} {Nature Physics}\ }\textbf {\bibinfo {volume} {10}},\ \bibinfo
  {pages} {116} (\bibinfo {year} {2014})}\BibitemShut {NoStop}%
\bibitem [{\citenamefont {Eigen}\ \emph {et~al.}(2017)\citenamefont {Eigen},
  \citenamefont {Glidden}, \citenamefont {Lopes}, \citenamefont {Navon},
  \citenamefont {Hadzibabic},\ and\ \citenamefont {Smith}}]{dynamics2}%
  \BibitemOpen
  \bibfield  {author} {\bibinfo {author} {\bibfnamefont {C.}~\bibnamefont
  {Eigen}}, \bibinfo {author} {\bibfnamefont {J.~A.~P.}\ \bibnamefont
  {Glidden}}, \bibinfo {author} {\bibfnamefont {R.}~\bibnamefont {Lopes}},
  \bibinfo {author} {\bibfnamefont {N.}~\bibnamefont {Navon}}, \bibinfo
  {author} {\bibfnamefont {Z.}~\bibnamefont {Hadzibabic}}, \ and\ \bibinfo
  {author} {\bibfnamefont {R.~P.}\ \bibnamefont {Smith}},\ }\bibfield  {title}
  {\bibinfo {title} {Universal scaling laws in the dynamics of a homogeneous
  unitary bose gas},\ }\href {\doibase/10.1103/PhysRevLett.119.250404}
  {\bibfield  {journal} {\bibinfo  {journal} {Phys. Rev. Lett.}\ }\textbf
  {\bibinfo {volume} {119}},\ \bibinfo {pages} {250404} (\bibinfo {year}
  {2017})}\BibitemShut {NoStop}%
\bibitem [{\citenamefont {Pr{\"u}fer}\ \emph {et~al.}(2018)\citenamefont
  {Pr{\"u}fer}, \citenamefont {Kunkel}, \citenamefont {Strobel}, \citenamefont
  {Lannig}, \citenamefont {Linnemann}, \citenamefont {Schmied}, \citenamefont
  {Berges}, \citenamefont {Gasenzer},\ and\ \citenamefont
  {Oberthaler}}]{dynamics3}%
  \BibitemOpen
  \bibfield  {author} {\bibinfo {author} {\bibfnamefont {M.}~\bibnamefont
  {Pr{\"u}fer}}, \bibinfo {author} {\bibfnamefont {P.}~\bibnamefont {Kunkel}},
  \bibinfo {author} {\bibfnamefont {H.}~\bibnamefont {Strobel}}, \bibinfo
  {author} {\bibfnamefont {S.}~\bibnamefont {Lannig}}, \bibinfo {author}
  {\bibfnamefont {D.}~\bibnamefont {Linnemann}}, \bibinfo {author}
  {\bibfnamefont {C.-M.}\ \bibnamefont {Schmied}}, \bibinfo {author}
  {\bibfnamefont {J.}~\bibnamefont {Berges}}, \bibinfo {author} {\bibfnamefont
  {T.}~\bibnamefont {Gasenzer}}, \ and\ \bibinfo {author} {\bibfnamefont
  {M.~K.}\ \bibnamefont {Oberthaler}},\ }\bibfield  {title} {\bibinfo {title}
  {Observation of universal dynamics in a spinor bose gas far from
  equilibrium},\ }\href {\doibase/10.1038/s41586-018-0659-0} {\bibfield
  {journal} {\bibinfo  {journal} {Nature}\ }\textbf {\bibinfo {volume} {563}},\
  \bibinfo {pages} {217} (\bibinfo {year} {2018})}\BibitemShut {NoStop}%
\bibitem [{\citenamefont {Erne}\ \emph {et~al.}(2018)\citenamefont {Erne},
  \citenamefont {B{\"u}cker}, \citenamefont {Gasenzer}, \citenamefont
  {Berges},\ and\ \citenamefont {Schmiedmayer}}]{dynamics4}%
  \BibitemOpen
  \bibfield  {author} {\bibinfo {author} {\bibfnamefont {S.}~\bibnamefont
  {Erne}}, \bibinfo {author} {\bibfnamefont {R.}~\bibnamefont {B{\"u}cker}},
  \bibinfo {author} {\bibfnamefont {T.}~\bibnamefont {Gasenzer}}, \bibinfo
  {author} {\bibfnamefont {J.}~\bibnamefont {Berges}}, \ and\ \bibinfo {author}
  {\bibfnamefont {J.}~\bibnamefont {Schmiedmayer}},\ }\bibfield  {title}
  {\bibinfo {title} {Universal dynamics in an isolated one-dimensional bose gas
  far from equilibrium},\ }\href {\doibase/10.1038/s41586-018-0667-0}
  {\bibfield  {journal} {\bibinfo  {journal} {Nature}\ }\textbf {\bibinfo
  {volume} {563}},\ \bibinfo {pages} {225} (\bibinfo {year}
  {2018})}\BibitemShut {NoStop}%
\bibitem [{\citenamefont {Eigen}\ \emph {et~al.}(2018)\citenamefont {Eigen},
  \citenamefont {Gliden}, \citenamefont {Lopes}, \citenamefont {Cornel},
  \citenamefont {Smith},\ and\ \citenamefont {Hadzibabic}}]{Bosequenchexp}%
  \BibitemOpen
  \bibfield  {author} {\bibinfo {author} {\bibfnamefont {C.}~\bibnamefont
  {Eigen}}, \bibinfo {author} {\bibfnamefont {J.~A.~P.}\ \bibnamefont
  {Gliden}}, \bibinfo {author} {\bibfnamefont {R.}~\bibnamefont {Lopes}},
  \bibinfo {author} {\bibfnamefont {E.~A.}\ \bibnamefont {Cornel}}, \bibinfo
  {author} {\bibfnamefont {R.~P.}\ \bibnamefont {Smith}}, \ and\ \bibinfo
  {author} {\bibfnamefont {Z.}~\bibnamefont {Hadzibabic}},\ }\bibfield  {title}
  {\bibinfo {title} {Universal prethermal dynamics of bose gases quenched to
  unitarity},\ }\href {\doibase/10.1038/s41586-018-0674-1} {\bibfield
  {journal} {\bibinfo  {journal} {Natrue}\ }\textbf {\bibinfo {volume} {563}},\
  \bibinfo {pages} {221} (\bibinfo {year} {2018})}\BibitemShut {NoStop}%
\bibitem [{\citenamefont {Saint-Jalm}\ \emph {et~al.}(2019)\citenamefont
  {Saint-Jalm}, \citenamefont {Castilho}, \citenamefont {Le~Cerf},
  \citenamefont {Bakkali-Hassani}, \citenamefont {Ville}, \citenamefont
  {Nascimbene}, \citenamefont {Beugnon},\ and\ \citenamefont
  {Dalibard}}]{dynamics5}%
  \BibitemOpen
  \bibfield  {author} {\bibinfo {author} {\bibfnamefont {R.}~\bibnamefont
  {Saint-Jalm}}, \bibinfo {author} {\bibfnamefont {P.~C.~M.}\ \bibnamefont
  {Castilho}}, \bibinfo {author} {\bibfnamefont {E.}~\bibnamefont {Le~Cerf}},
  \bibinfo {author} {\bibfnamefont {B.}~\bibnamefont {Bakkali-Hassani}},
  \bibinfo {author} {\bibfnamefont {J.-L.}\ \bibnamefont {Ville}}, \bibinfo
  {author} {\bibfnamefont {S.}~\bibnamefont {Nascimbene}}, \bibinfo {author}
  {\bibfnamefont {J.}~\bibnamefont {Beugnon}}, \ and\ \bibinfo {author}
  {\bibfnamefont {J.}~\bibnamefont {Dalibard}},\ }\bibfield  {title} {\bibinfo
  {title} {Dynamical symmetry and breathers in a two-dimensional bose gas},\
  }\href {\doibase/10.1103/PhysRevX.9.021035} {\bibfield  {journal} {\bibinfo
  {journal} {Phys. Rev. X}\ }\textbf {\bibinfo {volume} {9}},\ \bibinfo {pages}
  {021035} (\bibinfo {year} {2019})}\BibitemShut {NoStop}%
\bibitem [{\citenamefont {Deng}\ \emph {et~al.}(2016)\citenamefont {Deng},
  \citenamefont {Shi}, \citenamefont {Diao}, \citenamefont {Yu}, \citenamefont
  {Zhai}, \citenamefont {Qi},\ and\ \citenamefont {Wu}}]{dynamics6}%
  \BibitemOpen
  \bibfield  {author} {\bibinfo {author} {\bibfnamefont {S.}~\bibnamefont
  {Deng}}, \bibinfo {author} {\bibfnamefont {Z.-Y.}\ \bibnamefont {Shi}},
  \bibinfo {author} {\bibfnamefont {P.}~\bibnamefont {Diao}}, \bibinfo {author}
  {\bibfnamefont {Q.}~\bibnamefont {Yu}}, \bibinfo {author} {\bibfnamefont
  {H.}~\bibnamefont {Zhai}}, \bibinfo {author} {\bibfnamefont {R.}~\bibnamefont
  {Qi}}, \ and\ \bibinfo {author} {\bibfnamefont {H.}~\bibnamefont {Wu}},\
  }\bibfield  {title} {\bibinfo {title} {Observation of the efimovian expansion
  in scale-invariant fermi gases},\ }\href {\doibase/10.1126/science.aaf0666}
  {\bibfield  {journal} {\bibinfo  {journal} {Science}\ }\textbf {\bibinfo
  {volume} {353}},\ \bibinfo {pages} {371} (\bibinfo {year}
  {2016})}\BibitemShut {NoStop}%
\bibitem [{\citenamefont {Hu}\ \emph {et~al.}(2019)\citenamefont {Hu},
  \citenamefont {Feng}, \citenamefont {Zhang},\ and\ \citenamefont
  {Chin}}]{ChengChindynamics}%
  \BibitemOpen
  \bibfield  {author} {\bibinfo {author} {\bibfnamefont {J.}~\bibnamefont
  {Hu}}, \bibinfo {author} {\bibfnamefont {L.}~\bibnamefont {Feng}}, \bibinfo
  {author} {\bibfnamefont {Z.}~\bibnamefont {Zhang}}, \ and\ \bibinfo {author}
  {\bibfnamefont {C.}~\bibnamefont {Chin}},\ }\bibfield  {title} {\bibinfo
  {title} {Quantum simulation of unruh radiation},\ }\href
  {\doibase/10.1038/s41567-019-0537-1} {\bibfield  {journal} {\bibinfo
  {journal} {Nature Physics}\ }\textbf {\bibinfo {volume} {15}},\ \bibinfo
  {pages} {785} (\bibinfo {year} {2019})}\BibitemShut {NoStop}%
\bibitem [{\citenamefont {Lv}\ \emph {et~al.}(2020)\citenamefont {Lv},
  \citenamefont {Zhang},\ and\ \citenamefont {Zhou}}]{SU11echo}%
  \BibitemOpen
  \bibfield  {author} {\bibinfo {author} {\bibfnamefont {C.}~\bibnamefont
  {Lv}}, \bibinfo {author} {\bibfnamefont {R.}~\bibnamefont {Zhang}}, \ and\
  \bibinfo {author} {\bibfnamefont {Q.}~\bibnamefont {Zhou}},\ }\bibfield
  {title} {\bibinfo {title} {$su(1,1)$ echoes for breathers in quantum gases},\
  }\href {\doibase/10.1103/PhysRevLett.125.253002} {\bibfield  {journal}
  {\bibinfo  {journal} {Phys. Rev. Lett.}\ }\textbf {\bibinfo {volume} {125}},\
  \bibinfo {pages} {253002} (\bibinfo {year} {2020})}\BibitemShut {NoStop}%
\bibitem [{\citenamefont {Chen}\ \emph {et~al.}(2020)\citenamefont {Chen},
  \citenamefont {Zhang}, \citenamefont {Zheng}, \citenamefont {Wu},\ and\
  \citenamefont {Zhai}}]{MBE}%
  \BibitemOpen
  \bibfield  {author} {\bibinfo {author} {\bibfnamefont {Y.-Y.}\ \bibnamefont
  {Chen}}, \bibinfo {author} {\bibfnamefont {P.}~\bibnamefont {Zhang}},
  \bibinfo {author} {\bibfnamefont {W.}~\bibnamefont {Zheng}}, \bibinfo
  {author} {\bibfnamefont {Z.}~\bibnamefont {Wu}}, \ and\ \bibinfo {author}
  {\bibfnamefont {H.}~\bibnamefont {Zhai}},\ }\bibfield  {title} {\bibinfo
  {title} {Many-body echo},\ }\href {\doibase/10.1103/PhysRevA.102.011301}
  {\bibfield  {journal} {\bibinfo  {journal} {Phys. Rev. A}\ }\textbf {\bibinfo
  {volume} {102}},\ \bibinfo {pages} {011301} (\bibinfo {year}
  {2020})}\BibitemShut {NoStop}%
\bibitem [{\citenamefont {Abanin}\ \emph {et~al.}(2019)\citenamefont {Abanin},
  \citenamefont {Altman}, \citenamefont {Bloch},\ and\ \citenamefont
  {Serbyn}}]{MBLdynamics}%
  \BibitemOpen
  \bibfield  {author} {\bibinfo {author} {\bibfnamefont {D.~A.}\ \bibnamefont
  {Abanin}}, \bibinfo {author} {\bibfnamefont {E.}~\bibnamefont {Altman}},
  \bibinfo {author} {\bibfnamefont {I.}~\bibnamefont {Bloch}}, \ and\ \bibinfo
  {author} {\bibfnamefont {M.}~\bibnamefont {Serbyn}},\ }\bibfield  {title}
  {\bibinfo {title} {Colloquium: Many-body localization, thermalization, and
  entanglement},\ }\href {\doibase/10.1103/RevModPhys.91.021001} {\bibfield
  {journal} {\bibinfo  {journal} {Rev. Mod. Phys.}\ }\textbf {\bibinfo {volume}
  {91}},\ \bibinfo {pages} {021001} (\bibinfo {year} {2019})}\BibitemShut
  {NoStop}%
\bibitem [{\citenamefont {Wang}\ \emph {et~al.}(2017)\citenamefont {Wang},
  \citenamefont {Zhang}, \citenamefont {Chen}, \citenamefont {Yu},\ and\
  \citenamefont {Zhai}}]{topo1}%
  \BibitemOpen
  \bibfield  {author} {\bibinfo {author} {\bibfnamefont {C.}~\bibnamefont
  {Wang}}, \bibinfo {author} {\bibfnamefont {P.}~\bibnamefont {Zhang}},
  \bibinfo {author} {\bibfnamefont {X.}~\bibnamefont {Chen}}, \bibinfo {author}
  {\bibfnamefont {J.}~\bibnamefont {Yu}}, \ and\ \bibinfo {author}
  {\bibfnamefont {H.}~\bibnamefont {Zhai}},\ }\bibfield  {title} {\bibinfo
  {title} {Scheme to measure the topological number of a chern insulator from
  quench dynamics},\ }\href {\doibase/10.1103/PhysRevLett.118.185701}
  {\bibfield  {journal} {\bibinfo  {journal} {Phys. Rev. Lett.}\ }\textbf
  {\bibinfo {volume} {118}},\ \bibinfo {pages} {185701} (\bibinfo {year}
  {2017})}\BibitemShut {NoStop}%
\bibitem [{\citenamefont {Tarnowski}\ \emph {et~al.}(2019)\citenamefont
  {Tarnowski}, \citenamefont {{\"U}nal}, \citenamefont {Fl{\"a}schner},
  \citenamefont {Rem}, \citenamefont {Eckardt}, \citenamefont {Sengstock},\
  and\ \citenamefont {Weitenberg}}]{topo2}%
  \BibitemOpen
  \bibfield  {author} {\bibinfo {author} {\bibfnamefont {M.}~\bibnamefont
  {Tarnowski}}, \bibinfo {author} {\bibfnamefont {F.~N.}\ \bibnamefont
  {{\"U}nal}}, \bibinfo {author} {\bibfnamefont {N.}~\bibnamefont
  {Fl{\"a}schner}}, \bibinfo {author} {\bibfnamefont {B.~S.}\ \bibnamefont
  {Rem}}, \bibinfo {author} {\bibfnamefont {A.}~\bibnamefont {Eckardt}},
  \bibinfo {author} {\bibfnamefont {K.}~\bibnamefont {Sengstock}}, \ and\
  \bibinfo {author} {\bibfnamefont {C.}~\bibnamefont {Weitenberg}},\ }\bibfield
   {title} {\bibinfo {title} {Measuring topology from dynamics by obtaining the
  chern number from a linking number},\ }\href
  {\doibase/10.1038/s41467-019-09668-y} {\bibfield  {journal} {\bibinfo
  {journal} {Nature Communications}\ }\textbf {\bibinfo {volume} {10}},\
  \bibinfo {pages} {1728} (\bibinfo {year} {2019})}\BibitemShut {NoStop}%
\bibitem [{\citenamefont {Sun}\ \emph {et~al.}(2018)\citenamefont {Sun},
  \citenamefont {Yi}, \citenamefont {Wang}, \citenamefont {Zhang},
  \citenamefont {Sanders}, \citenamefont {Xu}, \citenamefont {Wang},
  \citenamefont {Schmiedmayer}, \citenamefont {Deng}, \citenamefont {Liu},
  \citenamefont {Chen},\ and\ \citenamefont {Pan}}]{topo3}%
  \BibitemOpen
  \bibfield  {author} {\bibinfo {author} {\bibfnamefont {W.}~\bibnamefont
  {Sun}}, \bibinfo {author} {\bibfnamefont {C.-R.}\ \bibnamefont {Yi}},
  \bibinfo {author} {\bibfnamefont {B.-Z.}\ \bibnamefont {Wang}}, \bibinfo
  {author} {\bibfnamefont {W.-W.}\ \bibnamefont {Zhang}}, \bibinfo {author}
  {\bibfnamefont {B.~C.}\ \bibnamefont {Sanders}}, \bibinfo {author}
  {\bibfnamefont {X.-T.}\ \bibnamefont {Xu}}, \bibinfo {author} {\bibfnamefont
  {Z.-Y.}\ \bibnamefont {Wang}}, \bibinfo {author} {\bibfnamefont
  {J.}~\bibnamefont {Schmiedmayer}}, \bibinfo {author} {\bibfnamefont
  {Y.}~\bibnamefont {Deng}}, \bibinfo {author} {\bibfnamefont {X.-J.}\
  \bibnamefont {Liu}}, \bibinfo {author} {\bibfnamefont {S.}~\bibnamefont
  {Chen}}, \ and\ \bibinfo {author} {\bibfnamefont {J.-W.}\ \bibnamefont
  {Pan}},\ }\bibfield  {title} {\bibinfo {title} {Uncover topology by quantum
  quench dynamics},\ }\href {\doibase/10.1103/PhysRevLett.121.250403}
  {\bibfield  {journal} {\bibinfo  {journal} {Phys. Rev. Lett.}\ }\textbf
  {\bibinfo {volume} {121}},\ \bibinfo {pages} {250403} (\bibinfo {year}
  {2018})}\BibitemShut {NoStop}%
\bibitem [{\citenamefont {Gao}\ \emph {et~al.}(2019)\citenamefont {Gao},
  \citenamefont {Zhai},\ and\ \citenamefont {Shi}}]{zheyudynamics}%
  \BibitemOpen
  \bibfield  {author} {\bibinfo {author} {\bibfnamefont {C.}~\bibnamefont
  {Gao}}, \bibinfo {author} {\bibfnamefont {H.}~\bibnamefont {Zhai}}, \ and\
  \bibinfo {author} {\bibfnamefont {Z.-Y.}\ \bibnamefont {Shi}},\ }\bibfield
  {title} {\bibinfo {title} {Dynamical fractal in quantum gases with discrete
  scaling symmetry},\ }\href {\doibase/10.1103/PhysRevLett.122.230402}
  {\bibfield  {journal} {\bibinfo  {journal} {Phys. Rev. Lett.}\ }\textbf
  {\bibinfo {volume} {122}},\ \bibinfo {pages} {230402} (\bibinfo {year}
  {2019})}\BibitemShut {NoStop}%
\bibitem [{\citenamefont {Gao}\ \emph {et~al.}(2020)\citenamefont {Gao},
  \citenamefont {Sun}, \citenamefont {Zhang},\ and\ \citenamefont
  {Zhai}}]{Chaodynamics}%
  \BibitemOpen
  \bibfield  {author} {\bibinfo {author} {\bibfnamefont {C.}~\bibnamefont
  {Gao}}, \bibinfo {author} {\bibfnamefont {M.}~\bibnamefont {Sun}}, \bibinfo
  {author} {\bibfnamefont {P.}~\bibnamefont {Zhang}}, \ and\ \bibinfo {author}
  {\bibfnamefont {H.}~\bibnamefont {Zhai}},\ }\bibfield  {title} {\bibinfo
  {title} {Universal dynamics of a degenerate bose gas quenched to unitarity},\
  }\href {\doibase/10.1103/PhysRevLett.124.040403} {\bibfield  {journal}
  {\bibinfo  {journal} {Phys. Rev. Lett.}\ }\textbf {\bibinfo {volume} {124}},\
  \bibinfo {pages} {040403} (\bibinfo {year} {2020})}\BibitemShut {NoStop}%
\bibitem [{\citenamefont {Sun}\ \emph {et~al.}(2020)\citenamefont {Sun},
  \citenamefont {Zhang},\ and\ \citenamefont {Zhai}}]{SunPRL}%
  \BibitemOpen
  \bibfield  {author} {\bibinfo {author} {\bibfnamefont {M.}~\bibnamefont
  {Sun}}, \bibinfo {author} {\bibfnamefont {P.}~\bibnamefont {Zhang}}, \ and\
  \bibinfo {author} {\bibfnamefont {H.}~\bibnamefont {Zhai}},\ }\bibfield
  {title} {\bibinfo {title} {High temperature virial expansion to universal
  quench dynamics},\ }\href {\doibase/10.1103/PhysRevLett.125.110404}
  {\bibfield  {journal} {\bibinfo  {journal} {Phys. Rev. Lett.}\ }\textbf
  {\bibinfo {volume} {125}},\ \bibinfo {pages} {110404} (\bibinfo {year}
  {2020})}\BibitemShut {NoStop}%
\bibitem [{\citenamefont {Rupak}(2007)}]{Virial1}%
  \BibitemOpen
  \bibfield  {author} {\bibinfo {author} {\bibfnamefont {G.}~\bibnamefont
  {Rupak}},\ }\bibfield  {title} {\bibinfo {title} {Universality in a
  2-component fermi system at finite temperature},\ }\href
  {\doibase/10.1103/PhysRevLett.98.090403} {\bibfield  {journal} {\bibinfo
  {journal} {Phys. Rev. Lett.}\ }\textbf {\bibinfo {volume} {98}},\ \bibinfo
  {pages} {090403} (\bibinfo {year} {2007})}\BibitemShut {NoStop}%
\bibitem [{\citenamefont {Liu}\ \emph {et~al.}(2009)\citenamefont {Liu},
  \citenamefont {Hu},\ and\ \citenamefont {Drummond}}]{Virial2}%
  \BibitemOpen
  \bibfield  {author} {\bibinfo {author} {\bibfnamefont {X.-J.}\ \bibnamefont
  {Liu}}, \bibinfo {author} {\bibfnamefont {H.}~\bibnamefont {Hu}}, \ and\
  \bibinfo {author} {\bibfnamefont {P.~D.}\ \bibnamefont {Drummond}},\
  }\bibfield  {title} {\bibinfo {title} {Virial expansion for a strongly
  correlated fermi gas},\ }\href {\doibase/10.1103/PhysRevLett.102.160401}
  {\bibfield  {journal} {\bibinfo  {journal} {Phys. Rev. Lett.}\ }\textbf
  {\bibinfo {volume} {102}},\ \bibinfo {pages} {160401} (\bibinfo {year}
  {2009})}\BibitemShut {NoStop}%
\bibitem [{\citenamefont {Liu}\ \emph {et~al.}(2010)\citenamefont {Liu},
  \citenamefont {Hu},\ and\ \citenamefont {Drummond}}]{Virial3}%
  \BibitemOpen
  \bibfield  {author} {\bibinfo {author} {\bibfnamefont {X.-J.}\ \bibnamefont
  {Liu}}, \bibinfo {author} {\bibfnamefont {H.}~\bibnamefont {Hu}}, \ and\
  \bibinfo {author} {\bibfnamefont {P.~D.}\ \bibnamefont {Drummond}},\
  }\bibfield  {title} {\bibinfo {title} {Three attractively interacting
  fermions in a harmonic trap: Exact solution, ferromagnetism, and
  high-temperature thermodynamics},\ }\href
  {\doibase/10.1103/PhysRevA.82.023619} {\bibfield  {journal} {\bibinfo
  {journal} {Phys. Rev. A}\ }\textbf {\bibinfo {volume} {82}},\ \bibinfo
  {pages} {023619} (\bibinfo {year} {2010})}\BibitemShut {NoStop}%
\bibitem [{\citenamefont {Liu}(2013)}]{Virial4}%
  \BibitemOpen
  \bibfield  {author} {\bibinfo {author} {\bibfnamefont {X.-J.}\ \bibnamefont
  {Liu}},\ }\bibfield  {title} {\bibinfo {title} {Virial expansion for a
  strongly correlated fermi system and its application to ultracold atomic
  fermi gases},\ }\href
  {\doibase/https://doi.org/10.1016/j.physrep.2012.10.004} {\bibfield
  {journal} {\bibinfo  {journal} {Physics Reports}\ }\textbf {\bibinfo {volume}
  {524}},\ \bibinfo {pages} {37} (\bibinfo {year} {2013})},\ \bibinfo {note}
  {virial expansion for a strongly correlated Fermi system and its application
  to ultracold atomic Fermi gases}\BibitemShut {NoStop}%
\bibitem [{\citenamefont {Peng}\ \emph
  {et~al.}(2011{\natexlab{a}})\citenamefont {Peng}, \citenamefont {Li},
  \citenamefont {Drummond},\ and\ \citenamefont {Liu}}]{Virial5}%
  \BibitemOpen
  \bibfield  {author} {\bibinfo {author} {\bibfnamefont {S.-G.}\ \bibnamefont
  {Peng}}, \bibinfo {author} {\bibfnamefont {S.-Q.}\ \bibnamefont {Li}},
  \bibinfo {author} {\bibfnamefont {P.~D.}\ \bibnamefont {Drummond}}, \ and\
  \bibinfo {author} {\bibfnamefont {X.-J.}\ \bibnamefont {Liu}},\ }\bibfield
  {title} {\bibinfo {title} {High-temperature thermodynamics of strongly
  interacting $s$-wave and $p$-wave fermi gases in a harmonic trap},\ }\href
  {\doibase/10.1103/PhysRevA.83.063618} {\bibfield  {journal} {\bibinfo
  {journal} {Phys. Rev. A}\ }\textbf {\bibinfo {volume} {83}},\ \bibinfo
  {pages} {063618} (\bibinfo {year} {2011}{\natexlab{a}})}\BibitemShut
  {NoStop}%
\bibitem [{\citenamefont {Peng}\ \emph {et~al.}(2014)\citenamefont {Peng},
  \citenamefont {Zhao},\ and\ \citenamefont {Jiang}}]{Virial6}%
  \BibitemOpen
  \bibfield  {author} {\bibinfo {author} {\bibfnamefont {S.-G.}\ \bibnamefont
  {Peng}}, \bibinfo {author} {\bibfnamefont {S.-H.}\ \bibnamefont {Zhao}}, \
  and\ \bibinfo {author} {\bibfnamefont {K.}~\bibnamefont {Jiang}},\ }\bibfield
   {title} {\bibinfo {title} {Virial expansion of a harmonically trapped fermi
  gas across a narrow feshbach resonance},\ }\href
  {\doibase/10.1103/PhysRevA.89.013603} {\bibfield  {journal} {\bibinfo
  {journal} {Phys. Rev. A}\ }\textbf {\bibinfo {volume} {89}},\ \bibinfo
  {pages} {013603} (\bibinfo {year} {2014})}\BibitemShut {NoStop}%
\bibitem [{\citenamefont {Yan}\ and\ \citenamefont {Blume}(2015)}]{Virial7}%
  \BibitemOpen
  \bibfield  {author} {\bibinfo {author} {\bibfnamefont {Y.}~\bibnamefont
  {Yan}}\ and\ \bibinfo {author} {\bibfnamefont {D.}~\bibnamefont {Blume}},\
  }\bibfield  {title} {\bibinfo {title} {Energy and structural properties of
  $n$-boson clusters attached to three-body efimov states: Two-body zero-range
  interactions and the role of the three-body regulator},\ }\href
  {\doibase/10.1103/PhysRevA.92.033626} {\bibfield  {journal} {\bibinfo
  {journal} {Phys. Rev. A}\ }\textbf {\bibinfo {volume} {92}},\ \bibinfo
  {pages} {033626} (\bibinfo {year} {2015})}\BibitemShut {NoStop}%
\bibitem [{\citenamefont {Yan}\ and\ \citenamefont
  {Blume}(2016{\natexlab{a}})}]{Virial8}%
  \BibitemOpen
  \bibfield  {author} {\bibinfo {author} {\bibfnamefont {Y.}~\bibnamefont
  {Yan}}\ and\ \bibinfo {author} {\bibfnamefont {D.}~\bibnamefont {Blume}},\
  }\bibfield  {title} {\bibinfo {title} {Path-integral monte carlo
  determination of the fourth-order virial coefficient for a unitary
  two-component fermi gas with zero-range interactions},\ }\href
  {\doibase/10.1103/PhysRevLett.116.230401} {\bibfield  {journal} {\bibinfo
  {journal} {Phys. Rev. Lett.}\ }\textbf {\bibinfo {volume} {116}},\ \bibinfo
  {pages} {230401} (\bibinfo {year} {2016}{\natexlab{a}})}\BibitemShut
  {NoStop}%
\bibitem [{\citenamefont {Sun}\ and\ \citenamefont {Cui}(2017)}]{Virial9}%
  \BibitemOpen
  \bibfield  {author} {\bibinfo {author} {\bibfnamefont {M.}~\bibnamefont
  {Sun}}\ and\ \bibinfo {author} {\bibfnamefont {X.}~\bibnamefont {Cui}},\
  }\bibfield  {title} {\bibinfo {title} {Enhancing the efimov correlation in
  bose polarons with large mass imbalance},\ }\href
  {\doibase/10.1103/PhysRevA.96.022707} {\bibfield  {journal} {\bibinfo
  {journal} {Phys. Rev. A}\ }\textbf {\bibinfo {volume} {96}},\ \bibinfo
  {pages} {022707} (\bibinfo {year} {2017})}\BibitemShut {NoStop}%
\bibitem [{\citenamefont {Nishida}(2019)}]{Virial10}%
  \BibitemOpen
  \bibfield  {author} {\bibinfo {author} {\bibfnamefont {Y.}~\bibnamefont
  {Nishida}},\ }\bibfield  {title} {\bibinfo {title} {Viscosity spectral
  functions of resonating fermions in the quantum virial expansion},\ }\href
  {\doibase/https://doi.org/10.1016/j.aop.2019.167949} {\bibfield  {journal}
  {\bibinfo  {journal} {Annals of Physics}\ }\textbf {\bibinfo {volume}
  {410}},\ \bibinfo {pages} {167949} (\bibinfo {year} {2019})}\BibitemShut
  {NoStop}%
\bibitem [{\citenamefont {Hou}\ and\ \citenamefont
  {Drut}(2020{\natexlab{a}})}]{Virial11}%
  \BibitemOpen
  \bibfield  {author} {\bibinfo {author} {\bibfnamefont {Y.}~\bibnamefont
  {Hou}}\ and\ \bibinfo {author} {\bibfnamefont {J.~E.}\ \bibnamefont {Drut}},\
  }\bibfield  {title} {\bibinfo {title} {Virial expansion of attractively
  interacting fermi gases in one, two, and three dimensions, up to fifth
  order},\ }\href {\doibase/10.1103/PhysRevA.102.033319} {\bibfield  {journal}
  {\bibinfo  {journal} {Phys. Rev. A}\ }\textbf {\bibinfo {volume} {102}},\
  \bibinfo {pages} {033319} (\bibinfo {year} {2020}{\natexlab{a}})}\BibitemShut
  {NoStop}%
\bibitem [{\citenamefont {Hou}\ \emph {et~al.}(2021{\natexlab{a}})\citenamefont
  {Hou}, \citenamefont {Morrell}, \citenamefont {Czejdo},\ and\ \citenamefont
  {Drut}}]{Virial12}%
  \BibitemOpen
  \bibfield  {author} {\bibinfo {author} {\bibfnamefont {Y.}~\bibnamefont
  {Hou}}, \bibinfo {author} {\bibfnamefont {K.~J.}\ \bibnamefont {Morrell}},
  \bibinfo {author} {\bibfnamefont {A.~J.}\ \bibnamefont {Czejdo}}, \ and\
  \bibinfo {author} {\bibfnamefont {J.~E.}\ \bibnamefont {Drut}},\ }\bibfield
  {title} {\bibinfo {title} {Fourth- and fifth-order virial expansion of
  harmonically trapped fermions at unitarity},\ }\href
  {\doibase/10.1103/PhysRevResearch.3.033099} {\bibfield  {journal} {\bibinfo
  {journal} {Phys. Rev. Res.}\ }\textbf {\bibinfo {volume} {3}},\ \bibinfo
  {pages} {033099} (\bibinfo {year} {2021}{\natexlab{a}})}\BibitemShut
  {NoStop}%
\bibitem [{\citenamefont {Sun}\ \emph {et~al.}(2017)\citenamefont {Sun},
  \citenamefont {Zhai},\ and\ \citenamefont {Cui}}]{Virial13}%
  \BibitemOpen
  \bibfield  {author} {\bibinfo {author} {\bibfnamefont {M.}~\bibnamefont
  {Sun}}, \bibinfo {author} {\bibfnamefont {H.}~\bibnamefont {Zhai}}, \ and\
  \bibinfo {author} {\bibfnamefont {X.}~\bibnamefont {Cui}},\ }\bibfield
  {title} {\bibinfo {title} {Visualizing the efimov correlation in bose
  polarons},\ }\href {\doibase/10.1103/PhysRevLett.119.013401} {\bibfield
  {journal} {\bibinfo  {journal} {Phys. Rev. Lett.}\ }\textbf {\bibinfo
  {volume} {119}},\ \bibinfo {pages} {013401} (\bibinfo {year}
  {2017})}\BibitemShut {NoStop}%
\bibitem [{\citenamefont {Kaplan}\ and\ \citenamefont {Sun}(2011)}]{Virial14}%
  \BibitemOpen
  \bibfield  {author} {\bibinfo {author} {\bibfnamefont {D.~B.}\ \bibnamefont
  {Kaplan}}\ and\ \bibinfo {author} {\bibfnamefont {S.}~\bibnamefont {Sun}},\
  }\bibfield  {title} {\bibinfo {title} {New field-theoretic method for the
  virial expansion},\ }\href {\doibase/10.1103/PhysRevLett.107.030601}
  {\bibfield  {journal} {\bibinfo  {journal} {Phys. Rev. Lett.}\ }\textbf
  {\bibinfo {volume} {107}},\ \bibinfo {pages} {030601} (\bibinfo {year}
  {2011})}\BibitemShut {NoStop}%
\bibitem [{\citenamefont {Bourdel}\ \emph {et~al.}(2003)\citenamefont
  {Bourdel}, \citenamefont {Cubizolles}, \citenamefont {Khaykovich},
  \citenamefont {Magalh\~aes}, \citenamefont {Kokkelmans}, \citenamefont
  {Shlyapnikov},\ and\ \citenamefont {Salomon}}]{Virialexp1}%
  \BibitemOpen
  \bibfield  {author} {\bibinfo {author} {\bibfnamefont {T.}~\bibnamefont
  {Bourdel}}, \bibinfo {author} {\bibfnamefont {J.}~\bibnamefont {Cubizolles}},
  \bibinfo {author} {\bibfnamefont {L.}~\bibnamefont {Khaykovich}}, \bibinfo
  {author} {\bibfnamefont {K.~M.~F.}\ \bibnamefont {Magalh\~aes}}, \bibinfo
  {author} {\bibfnamefont {S.~J. J. M.~F.}\ \bibnamefont {Kokkelmans}},
  \bibinfo {author} {\bibfnamefont {G.~V.}\ \bibnamefont {Shlyapnikov}}, \ and\
  \bibinfo {author} {\bibfnamefont {C.}~\bibnamefont {Salomon}},\ }\bibfield
  {title} {\bibinfo {title} {Measurement of the interaction energy near a
  feshbach resonance in a $^{6}\mathrm{L}\mathrm{i}$ fermi gas},\ }\href
  {\doibase/10.1103/PhysRevLett.91.020402} {\bibfield  {journal} {\bibinfo
  {journal} {Phys. Rev. Lett.}\ }\textbf {\bibinfo {volume} {91}},\ \bibinfo
  {pages} {020402} (\bibinfo {year} {2003})}\BibitemShut {NoStop}%
\bibitem [{\citenamefont {Stewart}\ \emph {et~al.}(2008)\citenamefont
  {Stewart}, \citenamefont {Gaebler},\ and\ \citenamefont {Jin}}]{Virialexp2}%
  \BibitemOpen
  \bibfield  {author} {\bibinfo {author} {\bibfnamefont {J.~T.}\ \bibnamefont
  {Stewart}}, \bibinfo {author} {\bibfnamefont {J.~P.}\ \bibnamefont
  {Gaebler}}, \ and\ \bibinfo {author} {\bibfnamefont {D.~S.}\ \bibnamefont
  {Jin}},\ }\bibfield  {title} {\bibinfo {title} {Using photoemission
  spectroscopy to probe a strongly interacting fermi gas},\ }\href
  {\doibase/10.1038/nature07172} {\bibfield  {journal} {\bibinfo  {journal}
  {Nature}\ }\textbf {\bibinfo {volume} {454}},\ \bibinfo {pages} {744}
  (\bibinfo {year} {2008})}\BibitemShut {NoStop}%
\bibitem [{\citenamefont {Nascimb{\`e}ne}\ \emph {et~al.}(2010)\citenamefont
  {Nascimb{\`e}ne}, \citenamefont {Navon}, \citenamefont {Jiang}, \citenamefont
  {Chevy},\ and\ \citenamefont {Salomon}}]{Virialexp3}%
  \BibitemOpen
  \bibfield  {author} {\bibinfo {author} {\bibfnamefont {S.}~\bibnamefont
  {Nascimb{\`e}ne}}, \bibinfo {author} {\bibfnamefont {N.}~\bibnamefont
  {Navon}}, \bibinfo {author} {\bibfnamefont {K.~J.}\ \bibnamefont {Jiang}},
  \bibinfo {author} {\bibfnamefont {F.}~\bibnamefont {Chevy}}, \ and\ \bibinfo
  {author} {\bibfnamefont {C.}~\bibnamefont {Salomon}},\ }\bibfield  {title}
  {\bibinfo {title} {Exploring the thermodynamics of a universal fermi gas},\
  }\href {\doibase/10.1038/nature08814} {\bibfield  {journal} {\bibinfo
  {journal} {Nature}\ }\textbf {\bibinfo {volume} {463}},\ \bibinfo {pages}
  {1057} (\bibinfo {year} {2010})}\BibitemShut {NoStop}%
\bibitem [{\citenamefont {Kuhnle}\ \emph {et~al.}(2011)\citenamefont {Kuhnle},
  \citenamefont {Hoinka}, \citenamefont {Dyke}, \citenamefont {Hu},
  \citenamefont {Hannaford},\ and\ \citenamefont {Vale}}]{Virialexp4}%
  \BibitemOpen
  \bibfield  {author} {\bibinfo {author} {\bibfnamefont {E.~D.}\ \bibnamefont
  {Kuhnle}}, \bibinfo {author} {\bibfnamefont {S.}~\bibnamefont {Hoinka}},
  \bibinfo {author} {\bibfnamefont {P.}~\bibnamefont {Dyke}}, \bibinfo {author}
  {\bibfnamefont {H.}~\bibnamefont {Hu}}, \bibinfo {author} {\bibfnamefont
  {P.}~\bibnamefont {Hannaford}}, \ and\ \bibinfo {author} {\bibfnamefont
  {C.~J.}\ \bibnamefont {Vale}},\ }\bibfield  {title} {\bibinfo {title}
  {Temperature dependence of the universal contact parameter in a unitary fermi
  gas},\ }\href {\doibase/10.1103/PhysRevLett.106.170402} {\bibfield  {journal}
  {\bibinfo  {journal} {Phys. Rev. Lett.}\ }\textbf {\bibinfo {volume} {106}},\
  \bibinfo {pages} {170402} (\bibinfo {year} {2011})}\BibitemShut {NoStop}%
\bibitem [{\citenamefont {Feld}\ \emph {et~al.}(2011)\citenamefont {Feld},
  \citenamefont {Fr{\"o}hlich}, \citenamefont {Vogt}, \citenamefont
  {Koschorreck},\ and\ \citenamefont {K{\"o}hl}}]{Virialexp5}%
  \BibitemOpen
  \bibfield  {author} {\bibinfo {author} {\bibfnamefont {M.}~\bibnamefont
  {Feld}}, \bibinfo {author} {\bibfnamefont {B.}~\bibnamefont {Fr{\"o}hlich}},
  \bibinfo {author} {\bibfnamefont {E.}~\bibnamefont {Vogt}}, \bibinfo {author}
  {\bibfnamefont {M.}~\bibnamefont {Koschorreck}}, \ and\ \bibinfo {author}
  {\bibfnamefont {M.}~\bibnamefont {K{\"o}hl}},\ }\bibfield  {title} {\bibinfo
  {title} {Observation of a pairing pseudogap in a two-dimensional fermi gas},\
  }\href {\doibase/10.1038/nature10627} {\bibfield  {journal} {\bibinfo
  {journal} {Nature}\ }\textbf {\bibinfo {volume} {480}},\ \bibinfo {pages}
  {75} (\bibinfo {year} {2011})}\BibitemShut {NoStop}%
\bibitem [{\citenamefont {Ku}\ \emph {et~al.}(2012)\citenamefont {Ku},
  \citenamefont {Sommer}, \citenamefont {Cheuk},\ and\ \citenamefont
  {Zwierlein}}]{Virialexp6}%
  \BibitemOpen
  \bibfield  {author} {\bibinfo {author} {\bibfnamefont {M.~J.~H.}\
  \bibnamefont {Ku}}, \bibinfo {author} {\bibfnamefont {A.~T.}\ \bibnamefont
  {Sommer}}, \bibinfo {author} {\bibfnamefont {L.~W.}\ \bibnamefont {Cheuk}}, \
  and\ \bibinfo {author} {\bibfnamefont {M.~W.}\ \bibnamefont {Zwierlein}},\
  }\bibfield  {title} {\bibinfo {title} {Revealing the superfluid lambda
  transition in the universal thermodynamics of a unitary fermi gas},\ }\href
  {\doibase/10.1126/science.1214987} {\bibfield  {journal} {\bibinfo  {journal}
  {Science}\ }\textbf {\bibinfo {volume} {335}},\ \bibinfo {pages} {563}
  (\bibinfo {year} {2012})}\BibitemShut {NoStop}%
\bibitem [{\citenamefont {Mukherjee}\ \emph {et~al.}(2019)\citenamefont
  {Mukherjee}, \citenamefont {Patel}, \citenamefont {Yan}, \citenamefont
  {Fletcher}, \citenamefont {Struck},\ and\ \citenamefont
  {Zwierlein}}]{Virialexp7}%
  \BibitemOpen
  \bibfield  {author} {\bibinfo {author} {\bibfnamefont {B.}~\bibnamefont
  {Mukherjee}}, \bibinfo {author} {\bibfnamefont {P.~B.}\ \bibnamefont
  {Patel}}, \bibinfo {author} {\bibfnamefont {Z.}~\bibnamefont {Yan}}, \bibinfo
  {author} {\bibfnamefont {R.~J.}\ \bibnamefont {Fletcher}}, \bibinfo {author}
  {\bibfnamefont {J.}~\bibnamefont {Struck}}, \ and\ \bibinfo {author}
  {\bibfnamefont {M.~W.}\ \bibnamefont {Zwierlein}},\ }\bibfield  {title}
  {\bibinfo {title} {Spectral response and contact of the unitary fermi gas},\
  }\href {\doibase/10.1103/PhysRevLett.122.203402} {\bibfield  {journal}
  {\bibinfo  {journal} {Phys. Rev. Lett.}\ }\textbf {\bibinfo {volume} {122}},\
  \bibinfo {pages} {203402} (\bibinfo {year} {2019})}\BibitemShut {NoStop}%
\bibitem [{\citenamefont {Carcy}\ \emph {et~al.}(2019)\citenamefont {Carcy},
  \citenamefont {Hoinka}, \citenamefont {Lingham}, \citenamefont {Dyke},
  \citenamefont {Kuhn}, \citenamefont {Hu},\ and\ \citenamefont
  {Vale}}]{Virialexp8}%
  \BibitemOpen
  \bibfield  {author} {\bibinfo {author} {\bibfnamefont {C.}~\bibnamefont
  {Carcy}}, \bibinfo {author} {\bibfnamefont {S.}~\bibnamefont {Hoinka}},
  \bibinfo {author} {\bibfnamefont {M.~G.}\ \bibnamefont {Lingham}}, \bibinfo
  {author} {\bibfnamefont {P.}~\bibnamefont {Dyke}}, \bibinfo {author}
  {\bibfnamefont {C.~C.~N.}\ \bibnamefont {Kuhn}}, \bibinfo {author}
  {\bibfnamefont {H.}~\bibnamefont {Hu}}, \ and\ \bibinfo {author}
  {\bibfnamefont {C.~J.}\ \bibnamefont {Vale}},\ }\bibfield  {title} {\bibinfo
  {title} {Contact and sum rules in a near-uniform fermi gas at unitarity},\
  }\href {\doibase/10.1103/PhysRevLett.122.203401} {\bibfield  {journal}
  {\bibinfo  {journal} {Phys. Rev. Lett.}\ }\textbf {\bibinfo {volume} {122}},\
  \bibinfo {pages} {203401} (\bibinfo {year} {2019})}\BibitemShut {NoStop}%
\bibitem [{\citenamefont {Ho}\ and\ \citenamefont
  {Mueller}(2004)}]{wideF_virial}%
  \BibitemOpen
  \bibfield  {author} {\bibinfo {author} {\bibfnamefont {T.-L.}\ \bibnamefont
  {Ho}}\ and\ \bibinfo {author} {\bibfnamefont {E.~J.}\ \bibnamefont
  {Mueller}},\ }\bibfield  {title} {\bibinfo {title} {High temperature
  expansion applied to fermions near feshbach resonance},\ }\href
  {\doibase/10.1103/PhysRevLett.92.160404} {\bibfield  {journal} {\bibinfo
  {journal} {Phys. Rev. Lett.}\ }\textbf {\bibinfo {volume} {92}},\ \bibinfo
  {pages} {160404} (\bibinfo {year} {2004})}\BibitemShut {NoStop}%
\bibitem [{\citenamefont {Yan}\ and\ \citenamefont
  {Blume}(2016{\natexlab{b}})}]{wideF_virial1}%
  \BibitemOpen
  \bibfield  {author} {\bibinfo {author} {\bibfnamefont {Y.}~\bibnamefont
  {Yan}}\ and\ \bibinfo {author} {\bibfnamefont {D.}~\bibnamefont {Blume}},\
  }\bibfield  {title} {\bibinfo {title} {Path-integral monte carlo
  determination of the fourth-order virial coefficient for a unitary
  two-component fermi gas with zero-range interactions},\ }\href
  {\doibase/10.1103/PhysRevLett.116.230401} {\bibfield  {journal} {\bibinfo
  {journal} {Phys. Rev. Lett.}\ }\textbf {\bibinfo {volume} {116}},\ \bibinfo
  {pages} {230401} (\bibinfo {year} {2016}{\natexlab{b}})}\BibitemShut
  {NoStop}%
\bibitem [{\citenamefont {Yu}\ \emph {et~al.}(2015)\citenamefont {Yu},
  \citenamefont {Thywissen},\ and\ \citenamefont {Zhang}}]{wideF_virial2}%
  \BibitemOpen
  \bibfield  {author} {\bibinfo {author} {\bibfnamefont {Z.}~\bibnamefont
  {Yu}}, \bibinfo {author} {\bibfnamefont {J.~H.}\ \bibnamefont {Thywissen}}, \
  and\ \bibinfo {author} {\bibfnamefont {S.}~\bibnamefont {Zhang}},\ }\bibfield
   {title} {\bibinfo {title} {Universal relations for a fermi gas close to a
  $p$-wave interaction resonance},\ }\href
  {\doibase/10.1103/PhysRevLett.115.135304} {\bibfield  {journal} {\bibinfo
  {journal} {Phys. Rev. Lett.}\ }\textbf {\bibinfo {volume} {115}},\ \bibinfo
  {pages} {135304} (\bibinfo {year} {2015})}\BibitemShut {NoStop}%
\bibitem [{\citenamefont {Marcelino}\ \emph {et~al.}(2014)\citenamefont
  {Marcelino}, \citenamefont {Nicolai}, \citenamefont {Roditi},\ and\
  \citenamefont {LeClair}}]{wideF_virial3}%
  \BibitemOpen
  \bibfield  {author} {\bibinfo {author} {\bibfnamefont {E.}~\bibnamefont
  {Marcelino}}, \bibinfo {author} {\bibfnamefont {A.}~\bibnamefont {Nicolai}},
  \bibinfo {author} {\bibfnamefont {I.}~\bibnamefont {Roditi}}, \ and\ \bibinfo
  {author} {\bibfnamefont {A.}~\bibnamefont {LeClair}},\ }\bibfield  {title}
  {\bibinfo {title} {Virial coefficients for trapped bose and fermi gases
  beyond the unitary limit: An $s$-matrix approach},\ }\href
  {\doibase/10.1103/PhysRevA.90.053619} {\bibfield  {journal} {\bibinfo
  {journal} {Phys. Rev. A}\ }\textbf {\bibinfo {volume} {90}},\ \bibinfo
  {pages} {053619} (\bibinfo {year} {2014})}\BibitemShut {NoStop}%
\bibitem [{\citenamefont {Hou}\ and\ \citenamefont
  {Drut}(2020{\natexlab{b}})}]{wideF_virial4}%
  \BibitemOpen
  \bibfield  {author} {\bibinfo {author} {\bibfnamefont {Y.}~\bibnamefont
  {Hou}}\ and\ \bibinfo {author} {\bibfnamefont {J.~E.}\ \bibnamefont {Drut}},\
  }\bibfield  {title} {\bibinfo {title} {Virial expansion of attractively
  interacting fermi gases in one, two, and three dimensions, up to fifth
  order},\ }\href {\doibase/10.1103/PhysRevA.102.033319} {\bibfield  {journal}
  {\bibinfo  {journal} {Phys. Rev. A}\ }\textbf {\bibinfo {volume} {102}},\
  \bibinfo {pages} {033319} (\bibinfo {year} {2020}{\natexlab{b}})}\BibitemShut
  {NoStop}%
\bibitem [{\citenamefont {Ho}\ \emph {et~al.}(2012)\citenamefont {Ho},
  \citenamefont {Cui},\ and\ \citenamefont {Li}}]{narrowF_virial}%
  \BibitemOpen
  \bibfield  {author} {\bibinfo {author} {\bibfnamefont {T.-L.}\ \bibnamefont
  {Ho}}, \bibinfo {author} {\bibfnamefont {X.}~\bibnamefont {Cui}}, \ and\
  \bibinfo {author} {\bibfnamefont {W.}~\bibnamefont {Li}},\ }\bibfield
  {title} {\bibinfo {title} {Alternative route to strong interaction: Narrow
  feshbach resonance},\ }\href {\doibase/10.1103/PhysRevLett.108.250401}
  {\bibfield  {journal} {\bibinfo  {journal} {Phys. Rev. Lett.}\ }\textbf
  {\bibinfo {volume} {108}},\ \bibinfo {pages} {250401} (\bibinfo {year}
  {2012})}\BibitemShut {NoStop}%
\bibitem [{\citenamefont {Peng}\ \emph
  {et~al.}(2011{\natexlab{b}})\citenamefont {Peng}, \citenamefont {Liu},
  \citenamefont {Hu},\ and\ \citenamefont {Li}}]{narrowF_virial1}%
  \BibitemOpen
  \bibfield  {author} {\bibinfo {author} {\bibfnamefont {S.-G.}\ \bibnamefont
  {Peng}}, \bibinfo {author} {\bibfnamefont {X.-J.}\ \bibnamefont {Liu}},
  \bibinfo {author} {\bibfnamefont {H.}~\bibnamefont {Hu}}, \ and\ \bibinfo
  {author} {\bibfnamefont {S.-Q.}\ \bibnamefont {Li}},\ }\bibfield  {title}
  {\bibinfo {title} {Non-universal thermodynamics of a strongly interacting
  inhomogeneous fermi gas using the quantum virial expansion},\ }\href
  {\doibase/https://doi.org/10.1016/j.physleta.2011.06.045} {\bibfield
  {journal} {\bibinfo  {journal} {Physics Letters A}\ }\textbf {\bibinfo
  {volume} {375}},\ \bibinfo {pages} {2979} (\bibinfo {year}
  {2011}{\natexlab{b}})}\BibitemShut {NoStop}%
\bibitem [{\citenamefont {Hou}\ \emph {et~al.}(2021{\natexlab{b}})\citenamefont
  {Hou}, \citenamefont {Morrell}, \citenamefont {Czejdo},\ and\ \citenamefont
  {Drut}}]{narrowF_virial2}%
  \BibitemOpen
  \bibfield  {author} {\bibinfo {author} {\bibfnamefont {Y.}~\bibnamefont
  {Hou}}, \bibinfo {author} {\bibfnamefont {K.~J.}\ \bibnamefont {Morrell}},
  \bibinfo {author} {\bibfnamefont {A.~J.}\ \bibnamefont {Czejdo}}, \ and\
  \bibinfo {author} {\bibfnamefont {J.~E.}\ \bibnamefont {Drut}},\ }\bibfield
  {title} {\bibinfo {title} {Fourth- and fifth-order virial expansion of
  harmonically trapped fermions at unitarity},\ }\href
  {\doibase/10.1103/PhysRevResearch.3.033099} {\bibfield  {journal} {\bibinfo
  {journal} {Phys. Rev. Res.}\ }\textbf {\bibinfo {volume} {3}},\ \bibinfo
  {pages} {033099} (\bibinfo {year} {2021}{\natexlab{b}})}\BibitemShut
  {NoStop}%
\bibitem [{\citenamefont {Tajima}\ \emph {et~al.}(2021)\citenamefont {Tajima},
  \citenamefont {Tsutsui}, \citenamefont {Doi},\ and\ \citenamefont
  {Iida}}]{narrowF_virial3}%
  \BibitemOpen
  \bibfield  {author} {\bibinfo {author} {\bibfnamefont {H.}~\bibnamefont
  {Tajima}}, \bibinfo {author} {\bibfnamefont {S.}~\bibnamefont {Tsutsui}},
  \bibinfo {author} {\bibfnamefont {T.~M.}\ \bibnamefont {Doi}}, \ and\
  \bibinfo {author} {\bibfnamefont {K.}~\bibnamefont {Iida}},\ }\bibfield
  {title} {\bibinfo {title} {Unitary $p$-wave fermi gas in one dimension},\
  }\href {\doibase/10.1103/PhysRevA.104.023319} {\bibfield  {journal} {\bibinfo
   {journal} {Phys. Rev. A}\ }\textbf {\bibinfo {volume} {104}},\ \bibinfo
  {pages} {023319} (\bibinfo {year} {2021})}\BibitemShut {NoStop}%
\bibitem [{\citenamefont {Hofmann}(2020)}]{narrowF_virial4}%
  \BibitemOpen
  \bibfield  {author} {\bibinfo {author} {\bibfnamefont {J.}~\bibnamefont
  {Hofmann}},\ }\bibfield  {title} {\bibinfo {title} {High-temperature
  expansion of the viscosity in interacting quantum gases},\ }\href
  {\doibase/10.1103/PhysRevA.101.013620} {\bibfield  {journal} {\bibinfo
  {journal} {Phys. Rev. A}\ }\textbf {\bibinfo {volume} {101}},\ \bibinfo
  {pages} {013620} (\bibinfo {year} {2020})}\BibitemShut {NoStop}%
\bibitem [{\citenamefont {Gao}(1998)}]{QDT1}%
  \BibitemOpen
  \bibfield  {author} {\bibinfo {author} {\bibfnamefont {B.}~\bibnamefont
  {Gao}},\ }\bibfield  {title} {\bibinfo {title} {Quantum-defect theory of
  atomic collisions and molecular vibration spectra},\ }\href
  {\doibase/10.1103/PhysRevA.58.4222} {\bibfield  {journal} {\bibinfo
  {journal} {Phys. Rev. A}\ }\textbf {\bibinfo {volume} {58}},\ \bibinfo
  {pages} {4222} (\bibinfo {year} {1998})}\BibitemShut {NoStop}%
\bibitem [{\citenamefont {Gao}\ \emph {et~al.}(2005)\citenamefont {Gao},
  \citenamefont {Tiesinga}, \citenamefont {Williams},\ and\ \citenamefont
  {Julienne}}]{QDT2}%
  \BibitemOpen
  \bibfield  {author} {\bibinfo {author} {\bibfnamefont {B.}~\bibnamefont
  {Gao}}, \bibinfo {author} {\bibfnamefont {E.}~\bibnamefont {Tiesinga}},
  \bibinfo {author} {\bibfnamefont {C.~J.}\ \bibnamefont {Williams}}, \ and\
  \bibinfo {author} {\bibfnamefont {P.~S.}\ \bibnamefont {Julienne}},\
  }\bibfield  {title} {\bibinfo {title} {Multichannel quantum-defect theory for
  slow atomic collisions},\ }\href {\doibase/10.1103/PhysRevA.72.042719}
  {\bibfield  {journal} {\bibinfo  {journal} {Phys. Rev. A}\ }\textbf {\bibinfo
  {volume} {72}},\ \bibinfo {pages} {042719} (\bibinfo {year}
  {2005})}\BibitemShut {NoStop}%
\bibitem [{\citenamefont {Zhai}(2021)}]{zhai_book}%
  \BibitemOpen
  \bibfield  {author} {\bibinfo {author} {\bibfnamefont {H.}~\bibnamefont
  {Zhai}},\ }\href {\doibase/10.1017/9781108595216} {\emph {\bibinfo {title}
  {Ultracold Atomic Physics}}}\ (\bibinfo  {publisher} {Cambridge University
  Press},\ \bibinfo {year} {2021})\BibitemShut {NoStop}%
\end{thebibliography}

%

\onecolumngrid
\newpage
\setcounter{equation}{0}
\setcounter{figure}{0}
\setcounter{table}{0}
\makeatletter
\renewcommand{\theequation}{S\arabic{equation}}
\renewcommand{\thefigure}{S\arabic{figure}}
\renewcommand{\thetable}{S\arabic{table}}

\section{Supplementary material on ``Quench Dynamics of Thermal Bose Gases Across Wide and Narrow Feshbach''}
\subsection{1. Two-channel square well model }
\label{sec:app1}
We here solve the two-channel square model to show Feshbach resonances and use a spherical box potential $V$ with the interaction range of $r_0$ to describe the interaction between two atoms. We consider the $s$-wave scattering, so that the radial wave function is given by $\Psi=\chi/r$ with $\Psi$ being the wave function. $\chi$ satisfies the Schr\"{o}dinger equation,
\begin{align}
-\frac{1}{m}\frac{d^2\chi}{dr^2}{\mathbb I}+V(r)\chi=E\chi,
\end{align}
\begin{align}
V(r)=\begin{cases} 
\begin{bmatrix} -V_o & W \\ W & -V_c+\delta\mu_{B} B \end{bmatrix}, & \text{for } r<r_0; \\ \begin{bmatrix} 0 & 0 \\ 0 & \infty \end{bmatrix}, & \text{for } r>r_0.
\end{cases}
\end{align}
Here, $m$ is the mass of particles. $V_{c}$ and $V_{o}$ represent the closed channel $|c\rangle$ and open channel $|o\rangle$ potential, and $W$ is the inter-channel coupling strength. $\delta\mu_{B}$ is the magnetic momentum difference between closed and open channel. ${\mathbb I}$ denotes the $2\times2$ identity.

At the distance $r>r_0$, the wave function is written as
\begin{align}
\chi=A\sin(kr+\delta_0)|o\rangle. \label{eq:solu-out}
\end{align}
While, at a distance $r<r_0$, those two channels are coupled.
We define a new set of bases $|+\rangle$ and $|-\rangle$ such that the wave function and the Hamiltonian are diagonalized. Thus, the wave function can be rewritten as $\chi=\chi_+|+\rangle +\chi_-|-\rangle$, and the bases $|\pm\rangle$ are superposition of $|o\rangle$ and $|c\rangle$, 
\begin{align}
\begin{bmatrix} |+\rangle \\ |-\rangle \end{bmatrix}=\begin{bmatrix}\cos\theta & \sin\theta \\ -\sin\theta & \cos\theta \end{bmatrix}\begin{bmatrix} |o\rangle \\ |c\rangle \end{bmatrix}.
\end{align}
In the region $r<r_0$, the wave function satisfies the boundary condition $\chi(r=0)=0$, and the solution is given by
\begin{align}
\chi_\pm=C\sin\left(\sqrt{m(E-V_\pm)}r \right), \label{eq:solu-chi}
\end{align}
where $V_\pm=\frac{-V_c-V_o+\delta\mu_B B}{2}\pm\frac{1}{2}\sqrt{(V_c-V_o-\delta\mu_B B)^2+4W^2}$.

Considering the boundary condition at $r=r_0$, the close channel wave function vanishes and the open channel wave function keeps continuum. With the solution given in Eq.(\ref{eq:solu-out}) and (\ref{eq:solu-chi}), we obtian
\begin{align}
\frac{k}{\tan\delta_0}= \sqrt{m(E-V_+)}\cot\left(\sqrt{m(E-V_+)}r_0 \right)\cos^2\theta +\sqrt{m(E-V_-)}\cot\left(\sqrt{m(E-V_-)}r_0 \right)\sin^2\theta.\label{eq:phaseshift-B}
\end{align}

The relation between the scattering amplitude and the T-matrix reads
\begin{align}
f_{0}(k)=-\frac{1}{ik-k\cot\delta_0}=-\frac{m}{4\pi}T_{2}({\bf k}', {\bf k};E), \label{eq:s-amplitude}
\end{align}
thus we obtain the two-body scattering T-matrix with the phase shift given in Eq.(\ref{eq:phaseshift-B}).

Next, we consider the bound state energy. At the distance $r>r_0$, the wave function of bound state  is written as
\begin{align}
\chi=e^{-k_{\rm b}r}|o\rangle.
\end{align}
At the distance $r<r_0$, the same as Eq.(\ref{eq:solu-chi}), the solution is given by
\begin{align}
\chi_\pm=D\sin\left(\sqrt{m(-k_{\rm b}^2-V_\pm)}r \right).
\end{align}
By matching the boundary condition at $r=r_0$, we obtain
\begin{align}
-k_{\rm b}=&\sqrt{m(-k_{\rm b}^2-V_+)}\cot\left(\sqrt{m(-k_{\rm b}^2-V_+)}r_0 \right)\cos^2\theta +\sqrt{m(-k_{\rm b}^2-V_-)}\cot\left(\sqrt{m(-k_{\rm b}^2-V_-)}r_0 \right)\sin^2\theta. \label{eq:BE-wavefunction}
\end{align}
The solution $k_{\rm b}$ of Eq.(\ref{eq:BE-wavefunction}) should be a positive real.

\begin{figure}
\centering
\includegraphics[width=0.8\textwidth]{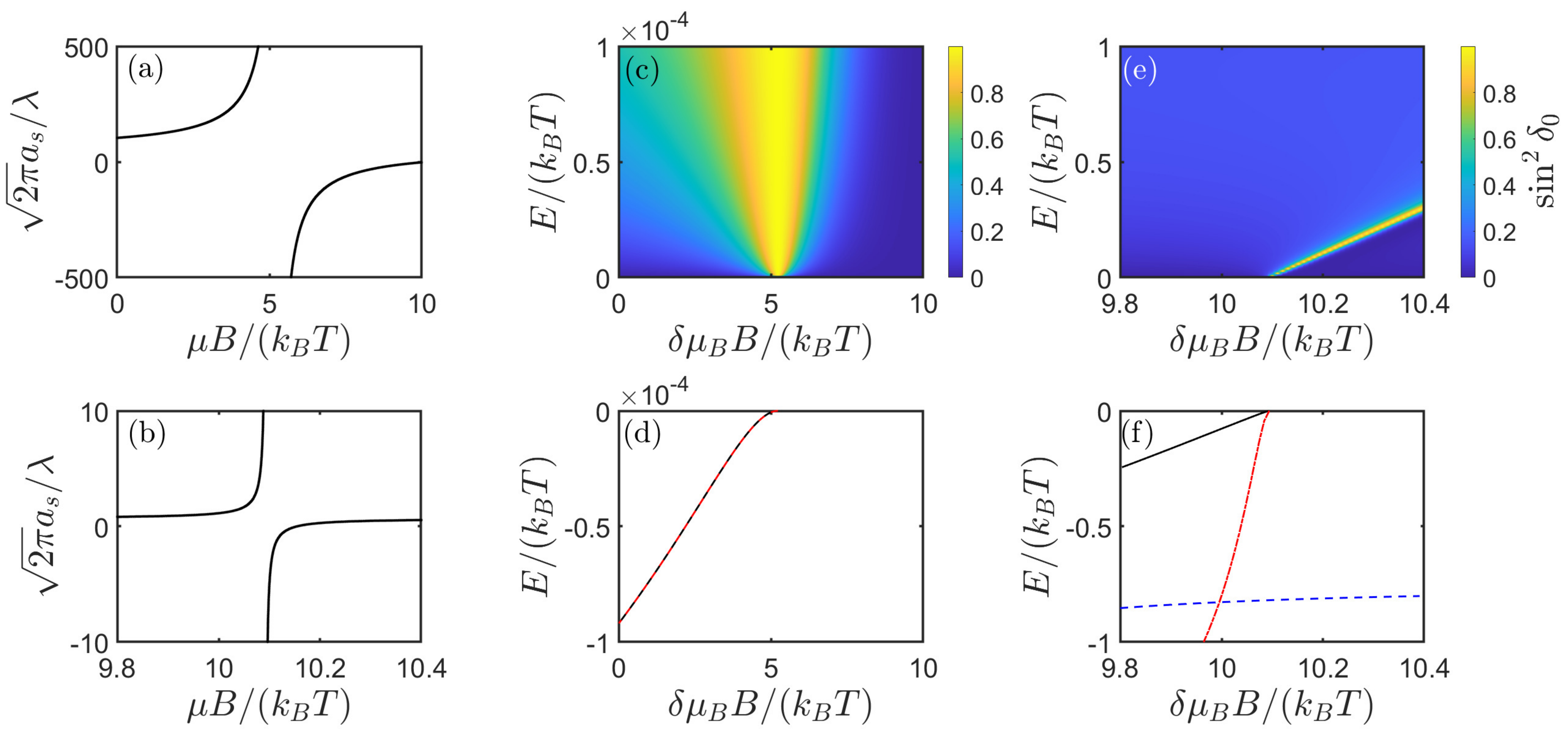}
\caption{ Feshbach resonances scattering length and the binding energy. (a) and (b) show the scattering length $a_s$ change with magnetic field strength $B$ under the wide and narrow resonances, respectively. (c) and (e) show $\sin^2\delta_0$ change with energy $E$. (d) and (f) show the energy of bound state, the soild line represent the binding energy given by Eq.(\ref{eq:BE-wavefunction}), the red dot-dashed line represent the binding energy given by $E=-\hbar^2/(ma_s^2)$, the blue dashed line in (f) represents another binding energy under narrow resonances.  Here, we choose $V_c=20 k_BT$, $V_o=2.5 k_BT$, and the coupling strength $W=0.5 k_BT$ as wide resonances in (a), (c) and (d), choose $V_c=20 k_BT$, $V_o=4.68 k_BT$, $W=0.35 k_BT$ as narrow resonances in (b), (e) and (f).} \label{fig:F-R}
\end{figure}

In order to distinguish wide and narrow resonances, we can define the parameter $s_{\rm res}$ as
\begin{align}
s_{\rm res}=\frac{a_{\rm bg}}{r_0}\frac{\delta\mu_B\Delta}{E_0},
\end{align}
where $a_{\rm bg}$ is the background scattering length, and $\Delta$ is the resonances width. $E_0=1/(mr_0^2)$. The background scattering length $a_{\rm bg}$ is determined by the open channel,
\begin{align}
a_{\rm bg}=\sqrt{-mV_o}\cot\left(\sqrt{-mV_o}r_0 \right).
\end{align}

Fig.~\ref{fig:F-R}(a) shows the scattering length $a_s$ of resonances with $s_{\rm res}=260$, it diverges at the position of resonance $\delta\mu_BB_{\rm res}=5.2 k_BT$. Fig.~\ref{fig:F-R}(c) shows a broad region of phase shift near the $B_{\rm res}$ where $\sin^2\delta_0\approx1$, and remains stable when the incoming energy increases. Fig.~\ref{fig:F-R}(d) shows the binding energy near threshold, The solid line represents the binding energy given by Eq.(\ref{eq:BE-wavefunction}), which is consistent well with the energy $E=-\hbar^2/(ma_s^2)$. In this resonances, we can also find another bound state energy, but those two energy difference is 4 or 5 orders of magnitude.

Fig.~\ref{fig:F-R}(b) shows the scattering length of narrow resonances with $s_{\rm res}=0.04$. Fig.~\ref{fig:F-R}(f) shows the binding energy near this resonances. First, we can see the energy given by the exact solution no longer matches the energy given by the single-channel model. This means that when dealing with a narrow resonance-related problems, we should use the two-channel model. Second, we obtain two bound states near the position of resonance, and one of the energies is neat threshold(solid line). The deeper energy is mainly contributed by the open channel, which is represented by the dashed line. Here, we choose $V_o=4.68  k_BT$, thus the bare bound state energy supported by the open channel is $E_{\rm b}=0.75 k_BT$, which is close to the deeper binding energy. 

\subsection{2. T-matrix of two-channel zero range model}
\begin{figure}
\centering
\includegraphics[width=0.8\textwidth]{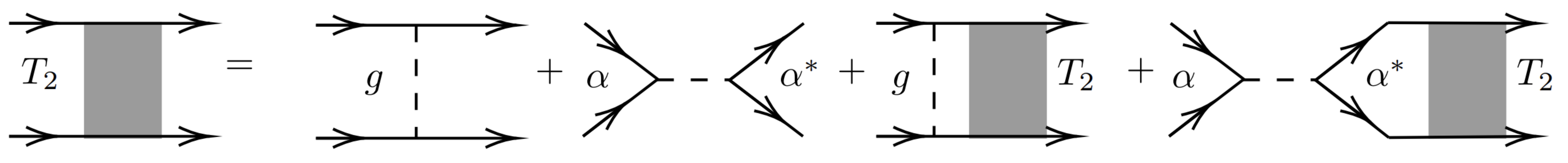}
\caption{ Ladder diagrams for two-body scattering T-matrix of the two-channel model.} \label{fig:Sp2}
\end{figure}

In this section, we derive the T-matrix of two-channel zero range model. With a contact potential, the second-quantized Hamiltonian in the momentum space can be written as  
\begin{align}
\mathcal{\hat H}=& \underset{\mathbf k \sigma}\sum \frac{\mathbf k^2}{2m}\hat \Psi_{\mathbf k \sigma}\hat\Psi_{\mathbf k \sigma}+\underset{\mathbf k}\sum\left(\frac{\mathbf k^2}{4m}+\nu\right)\hat b^\dag_{\mathbf k}\hat b_{\mathbf k}+\frac{g}{V}\underset{\mathbf k,\mathbf k_1,\mathbf k_2}\sum \Psi^\dag_{\frac{\mathbf k}{2}+\mathbf k_{1,\uparrow}}\Psi^\dag_{\frac{\mathbf k}{2}-\mathbf k_{1,\downarrow}}\Psi_{\frac{\mathbf k}{2}-\mathbf k_{2,\downarrow}}\Psi_{\frac{\mathbf k}{2}+\mathbf k_{2,\uparrow}} \\ \notag
&+\frac{\alpha}{\sqrt{V}}\underset{\mathbf k,\mathbf k_1}\sum \Psi^\dag_{\frac{\mathbf k}{2}+\mathbf k_{1,\uparrow}}\Psi^\dag_{\frac{\mathbf k}{2}-\mathbf k_{1,\downarrow}}\hat b_{\mathbf k}+\hat b^\dag_{\mathbf k}\Psi_{\frac{\mathbf k}{2}-\mathbf k_{1,\downarrow}}\Psi_{\frac{\mathbf k}{2}+\mathbf k_{1,\uparrow}}.
\end{align}
Here, $g$ is the bare interaction between open channel atoms themselves, $\Psi^\dag_{\sigma}$ and $\Psi_{\sigma}$ are the creation and annihilation operators for scattering states in the open channels and $V$ is the volume of the system. $\hat b^\dag$ and $\hat b$ are the creation and annihilation operators of the two-body bound state in the closed channel, and $\nu$ is the detuning of the molecular state in the closed channel. The last term denotes the conversion between the open channel scattering states and the closed channel molecular state, with the strength given by $\alpha$. The ladder diagram for the two-channel model is shown in Fig.~\ref{fig:Sp2}. The summation of the ladder diagram leads to the Schwinger-Dyson equation 

\begin{align}
T_2(E)=g+\frac{|\alpha|^2}{E-\nu}+\left(g+\frac{|\alpha|^2}{E-\nu}\right)\frac{1}{V}\underset{\mathbf k}\sum\frac{1}{E-\mathbf k^2/m}T_2(E),
\end{align}
which leads to
\begin{align}
T_2(E)=\frac{g+\frac{|\alpha|^2}{E-\nu}}{1-\left(g+\frac{|\alpha|^2}{E-\nu}\right)\frac{1}{V}\underset{\mathbf k}\sum\frac{1}{E-\mathbf k^2/m}}. \label{eq:ZeroRange-T}
\end{align}

Notice that the T-matrix given in Eq.(\ref{eq:ZeroRange-T}) is renormalizable. This two-body T-matrix should be related to the $s$-wave scattering amplitude, therefore, we have
\begin{align}
T_2(E=0)=\frac{4\pi}{m}f_0(k=0)=\frac{4\pi}{m}a_{\rm bg}\left(1-\frac{\Delta}{B-B_{\rm res}}\right).\label{eq:T2-E0}
\end{align}
We also define $\nu=\delta\mu_B(B-B_{\rm res})+\nu_p$.
When detuning the magnetic field away from the resonance position $|B-B_{\rm res}|\gg|\Delta|$, we have
\begin{align}
T_2(E=0)=\frac{4\pi}{m}a_{\rm bg}=\frac{1}{\frac{1}{g}+\frac{1}{V}\underset{\mathbf k}\sum\frac{1}{\mathbf k^2/m}}. 
\end{align}
Hence, we reach the renormalization identity that relates $g$ to physical quantity $a_{\rm bg}$,
\begin{align}
\frac{1}{g}=\frac{m}{4\pi a_{\rm bg}}-\Lambda,\label{g-re}
\end{align}
where $\Lambda$ denotes
\begin{align}
\Lambda=\frac{1}{V}\underset{\mathbf k}\sum\frac{1}{\mathbf k^2/m}.
\end{align}
By comparing Eq.(\ref{eq:ZeroRange-T}) with Eq.(\ref{eq:T2-E0}), and using the relation between $g$ and $a_{\rm bg}$, the remainder renormalization conditions can be given by
\begin{align}
\frac{1}{\alpha}=\left(1-\frac{4\pi a_{\rm bg}}{m}\Lambda \right)\sqrt{\frac{m}{4\pi a_{\rm bg}\delta\mu_B\Delta}},\label{alpha-re}
\end{align}
\begin{align}
\nu=\delta\mu_B(B-B_{\rm res})+\frac{\Lambda}{1-\frac{4\pi a_{\rm bg}}{m}\Lambda}\frac{4\pi a_{\rm bg}\delta\mu_B\Delta}{m}.\label{nu-re}
\end{align}

By substituting Eq.(\ref{g-re},\ref{alpha-re},\ref{nu-re}) into Eq.(\ref{eq:ZeroRange-T}), we have
\begin{align}
T_2=\frac{4\pi/m}{\frac{E-\delta\mu_B(B-B_{\rm res})}{a_{\rm bg}\left[E+\delta\mu_B\Delta-\delta\mu_B(B-B_{\rm res})\right]}-\sqrt{-mE}}.
\end{align}

\end{document}